\documentclass[10pt,journal,compsoc]{IEEEtran}
\usepackage[switch]{lineno}
\usepackage[l2tabu,orthodox]{nag}

\usepackage[T1]{fontenc}
\usepackage[utf8]{inputenc}
\usepackage[shrink=10]{microtype}

\usepackage{amsthm}
\usepackage{amssymb}
\usepackage{mathtools} %
\usepackage{textcomp}
\usepackage{siunitx} %
\usepackage{physics} %
\usepackage{stmaryrd}
\usepackage{scalerel}
\usepackage{bm}

\usepackage{ifthen}
\usepackage{float} %
\usepackage{hyperref}
\usepackage{cleveref}
\usepackage{csquotes}
\usepackage{parskip}
\usepackage{calc}
\usepackage[usenames,dvipsnames]{xcolor}

\usepackage[style=ieee,citestyle=numeric,sortcites=false,maxcitenames=1,mincitenames=1,uniquelist=false]{biblatex}

\usepackage{algorithm}
\usepackage{algorithmicx}
\usepackage{algpseudocode}
\usepackage{listingsutf8}

\usepackage{array}
\usepackage{booktabs}
\usepackage{multirow}

\usepackage{float}
\usepackage{wrapfig} %
\usepackage{graphicx}
\usepackage[percent]{overpic}
\usepackage[caption=false,font=footnotesize,labelfont=sf,textfont=sf]{subfig}
\usepackage{tikz}
\usetikzlibrary{decorations.pathmorphing,patterns,shadows,calc}

\definecolor{DanieleColor}{rgb}{0.56,0.34,0.62}
\definecolor{JeremieColor}{rgb}{0.82,0.33,0.00}
\definecolor{ZhenColor}{rgb}{0.55,0.35,0.05}

\newcommand{\daniele}[1]{{\leavevmode\color{DanieleColor} Daniele: #1 $\qed$}}
\newcommand{\DP}[1]{{\leavevmode\color{DanieleColor} Daniele: #1 $\qed$}}
\newcommand{\jeremie}[1]{{\leavevmode\color{JeremieColor} J\'{e}r\'{e}mie: #1 $\qed$}}
\newcommand{\zhen}[1]{{\leavevmode\color{ZhenColor} Zhen: #1 $\qed$}}

\newcommand{\toref}[1]{\textcolor{Red}{[ref:#1]}}
\newcommand{\tocite}[1]{\textcolor{Red}{[cite:#1]}}
\newcommand{\todo}[1]{\textcolor{Red}{TODO: #1 $\qed$}}
\newcommand{\warning}[1]{{\itshape\color{Red} #1}}

\newcommand{\note}[1]{{\itshape\color{blue} #1}}
\newcommand{\nothing}[1]{}

\newcommand{\rev}[1]{#1}

\providecommand{\finalversion}{1} %
\ifthenelse{\equal{\finalversion}{1}}
{
	\renewcommand{\DP}[1]{}
	\renewcommand{\daniele}[1]{}
	\renewcommand{\jeremie}[1]{}
	\renewcommand{\zhen}[1]{}
	\renewcommand{\toref}[1]{}
	\renewcommand{\tocite}[1]{}
	\renewcommand{\todo}[1]{}
	\renewcommand{\warning}[1]{}
	\renewcommand{\note}[1]{}
}
{}

\definecolor{forestgreen}{rgb}{0.13,0.54,0.13}
\definecolor{darkblue}{rgb}{0,0,.5}
\hypersetup{
	unicode=true,
	colorlinks=true,
	citecolor=forestgreen, %
	linkcolor=darkblue, %
	urlcolor=darkblue, %
}

\newcommand{\mcenter}[1]{\raisebox{-0.5\height}{#1}}

\newcommand{\tcenter}[1]{\raisebox{-1.0\height}{#1}}

\newcommand{\specialcell}[2][c]{%
	\begin{tabular}[#1]{@{}c@{}}#2\end{tabular}}

\newcommand{\fnurl}[2]{%
	\href{#1}{#2}\footnote{\url{#1}}%
}

\renewcommand{\paragraph}[1]{{\bfseries #1.}}
\renewcommand*{\paragraph}[1]{{\bfseries #1.}}

\algrenewcommand{\algorithmiccomment}[1]{{\footnotesize\color{forestgreen}\(\triangleright\) #1}}
\newcommand{\function}[1]{{\small\textsc{#1}}}

\newcommand{\lref}[1]{line~\ref{#1}}
\newcommand{\lrefs}[2]{lines~\ref{#1} and \ref{#2}}

\crefname{algocf}{alg.}{algs.}
\Crefname{algocf}{Algorithm}{Algorithms}

\crefname{appsec}{Appendix}{Appendices}

\newcommand{\inlinecode}[1]{\lstinline{#1}}

\newcommand*\patchAmsMathEnvironmentForLineno[1]{%
	\expandafter\let\csname old#1\expandafter\endcsname\csname #1\endcsname
	\expandafter\let\csname oldend#1\expandafter\endcsname\csname end#1\endcsname
	\renewenvironment{#1}%
	{\linenomath\csname old#1\endcsname}%
	{\csname oldend#1\endcsname\endlinenomath}}%
\newcommand*\patchBothAmsMathEnvironmentsForLineno[1]{%
	\patchAmsMathEnvironmentForLineno{#1}%
	\patchAmsMathEnvironmentForLineno{#1*}}%
\AtBeginDocument{%
	\patchBothAmsMathEnvironmentsForLineno{equation}%
	\patchBothAmsMathEnvironmentsForLineno{align}%
	\patchBothAmsMathEnvironmentsForLineno{flalign}%
	\patchBothAmsMathEnvironmentsForLineno{alignat}%
	\patchBothAmsMathEnvironmentsForLineno{gather}%
	\patchBothAmsMathEnvironmentsForLineno{multline}%
}

\let\originalleft\left
\let\originalright\right
\renewcommand{\left}{\mathopen{}\mathclose\bgroup\originalleft}
\renewcommand{\right}{\aftergroup\egroup\originalright}

\renewcommand{\geq}{\geqslant}
\renewcommand{\leq}{\leqslant}

\newcommand{\eqdef}{\stackrel{\text{\tiny def}}{=}}

\renewcommand{\O}{\mathcal{O}}

\newcommand{\reals}{\mathbb{R}}

\DeclareMathOperator\dist{dist}

\DeclarePairedDelimiter\irange{\llbracket}{\rrbracket} %
\newcommand{\compl}[1]{\overline{#1}} %

\DeclarePairedDelimiter\size{\lvert}{\rvert}
\renewcommand{\vec}[1]{{\bm{{#1}}}}

\DeclarePairedDelimiter\segment{[}{]}

\newcommand{\sIn}{^\vdash}
\newcommand{\sOut}{^\dashv}
\newcommand{\z}{\vec{z}}
\newcommand{\zIn}{\vec{z}\sIn}
\newcommand{\zOut}{\vec{z}\sOut}

\newcommand{\shape}{S}
\newcommand{\dilate}{{\mathcal D}}

\newcommand{\dilationRadius}{r}

\newcommand{\otherShape}{\shape'}
\newcommand{\shapeIn}{\shape_{\mathrm{In}}}
\newcommand{\shapeOut}{\shape_{\mathrm{Out}}}
\newcommand{\shapeMid}{\shape_{\mathrm{Mid}}}
\newcommand{\dilateEndpoints}{\dilate^\circ}
\newcommand{\dilateSegments}{\dilate^\Box}
\newcommand{\dilatePower}{\dilate^p}

\newcommand{\seeds}{\mathfrak{S}}
\newcommand{\seed}{\vec{\mathfrak{s}}}
\newcommand{\seedPoint}{\vec{s}}
\newcommand{\seedIn}{\vec{s}\sIn}
\newcommand{\seedOut}{\vec{s}\sOut}
\newcommand{\subseed}{\widetilde{\seed}}
\newcommand{\domain}{\Omega}
\newcommand{\voronoiCell}{\Omega}
\newcommand{\powerCell}{\Omega^{\mathrm{p}}}
\newcommand{\visdir}{\vec{v}}
\newcommand{\proj}[1]{\widetilde{#1}}
\newcommand{\weight}{w}

\newcommand{\Fwd}[1]{\overrightarrow{#1}}
\newcommand{\Rev}[1]{\overleftarrow{#1}}

\DeclareMathOperator\voronoi{Vor}
\newcommand{\Vor}[1]{\voronoi\left(#1\right)}

\newcommand{\mL}{\mathcal{L}}
\newcommand{\mQ}{\mathcal{Q}}

\newcommand{\point}{\vec{p}}
\newcommand{\otherPoint}{\vec{x}}

\begin{document}

\title{Half-Space Power Diagrams and Discrete Surface Offsets*}

\author{
	Zhen Chen, Daniele Panozzo, J\'{e}r\'{e}mie Dumas%
	\thanks{*: This work has been submitted to the IEEE for possible publication. Copyright may be transferred without notice, after which this version may no longer be accessible.}
}

\markboth{}{}

\IEEEtitleabstractindextext{%
	\begin{abstract}
		We present an efficient, trivially parallelizable algorithm to compute offset surfaces of shapes discretized using a \emph{dexel} data structure. %
Our algorithm is based on a two-stage sweeping procedure that is simple to implement and efficient, entirely avoiding volumetric distance field computations typical of existing methods. Our construction is based on properties of \emph{half-space} power diagrams, where each seed is only visible by a half-space, which were \rev{never} used before for the computation of surface offsets.

The primary application of our method is interactive modeling for digital fabrication. Our technique enables a user to interactively process high-resolution models. It is also useful in a plethora of other geometry processing tasks requiring fast, approximate offsets, such as topology optimization, collision detection, and skeleton extraction. We present experimental timings, comparisons with previous approaches, and provide a reference implementation in the supplemental material.

	\end{abstract}

}
\maketitle

\IEEEdisplaynontitleabstractindextext

\IEEEpeerreviewmaketitle

\section{Introduction}

\begin{figure}[bt!]
	\centering
	\captionsetup[subfigure]{labelformat=empty}

	\makebox[\linewidth][c]{
	\subfloat[Input Surface]{\includegraphics[height=5cm]{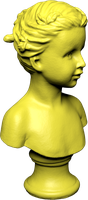}}
	\hfill
	\subfloat[Thick Shell]{\includegraphics[height=5cm]{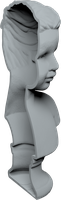}}
	\hfill
	\subfloat[3D Printed]{\includegraphics[height=5cm]{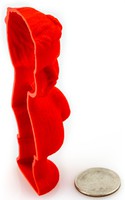}}
	}
	\caption{
		3D printed surface shell, computed with our approach.
		The shell is the Boolean difference of a dilation and an erosion of the input volume.
		The model uses a grid of $512^2$ dexels, and a dilation radius of 4 dexels.
	}
	\label{fig:teaser}
\end{figure}

Morphological operations, such as dilation and erosion, have numerous applications: They can be used to regularize shapes~\cite{Williams:2005:MMS}, to ensure robust designs in topology optimization~\cite{Sigmund:2007:MBB}, to perform collision detection~\cite{Teschner:2005:CDF}, or to compute image skeletons~\cite{Maragos:1986:MSR}. By combining these operations, it is possible to compute surface offsets. Offset surfaces are often used in digital fabrication applications~\cite{Livesu:2017:FDM}, to generate support structures~\cite{Hornus:2016:TPE}, to hollow object \rev{(\Cref{fig:teaser})}, to create molds, and to remove topological noise.

While offset surfaces can be computed exactly with Minkowski sums~\cite{Hachenberger:2007:EMS}, these operations can be slow, especially on large models. Recent approaches~\cite{Meng:2017:ECF} provide better results, but their performance is still insufficient for their use in interactive applications. Conversely, approximate algorithms which rely on a discrete re-sampling of the input volume, achieve efficiency by sacrificing accuracy in a controlled way~\cite{Wang:2013:GBO,Martinez:2015:CSO}. These methods are especially relevant in digital fabrication where resolutions are inherently limited by the machine fabrication tolerance, and using exact computation is unnecessary.

We propose a novel algorithm to compute offset surfaces on a solid object represented with a ray-based representation (ray-rep). A ray-rep, such as the \emph{dexel buffer}~\cite{VanHook:1986:RTS}, stores the intersections between a solid object and a set of parallel rays emanating from a uniform 2D grid: Each cell of the grid holds a list of \emph{intervals} bounding the solid (\Cref{fig:dexels}).
Ray-reps are appealing in the context of digital fabrication, since they allow us to compute morphological operations directly at the resolution of the machine.
CSG operations can be carried out directly in image space~\cite{Lefebvre:2013:IAG,Huang:2014:AFL}, at a fraction of the cost of their counterparts on meshes, for which fast and robust implementations are notoriously difficult~\cite{Zhou:2016:MAF}.
Another advantage is that implicit surfaces can be efficiently converted into ray-reps, avoiding explicit meshing. Examples in the literature can be found for Computer Numerical Control (CNC) milling applications~\cite{VanHook:1986:RTS}, modeling for additive manufacturing~\cite{Lefebvre:2013:IAG}, hollowing, or contouring~\cite{Livesu:2017:FDM}.
Our offset algorithm exploits the ray-rep representation, and it is both fast and \emph{accurate}, in the sense that it computes the \emph{exact} offset at the resolution used by the input dexel structure. %
We demonstrate experimentally that our algorithm scales well with the offset radius, in addition to being embarrassingly parallel and thus can fully exploit multi-core, shared-memory architectures.

\textbf{Contribution.} Our algorithm relies on a novel approach for computing half-plane Voronoi diagrams~\cite{Fan:2011:HPV}. We first describe this approach in 2D, demonstrating a simple and efficient sweeping algorithm to compute offsets in a 2D dexel structure.
We then extend our sweeping procedure to 3D by leveraging the separability of the Euclidean distance~\cite{Meijster:2002:AGA}.

In \Cref{sec:results}, we compare running times of our algorithm with existing approaches, and showcase applications of our approach in topological simplification and modeling for additive manufacturing.

\section{Related Work}
\label{sec:related-work}

\paragraph{Exact Mesh Offsets}
Dilated surfaces can be computed exactly for triangle meshes, by dilating every triangle and explicitly resolving the introduced self-intersections~\cite{Jung:2004:SIR,Campen:2010:EAR}. A dilated mesh can also be obtained by computing the Minkowski sum of the input triangle mesh and a sphere of the desired radius~\cite{Hachenberger:2007:EMS}. This approach, which is implemented robustly in CGAL~\cite{Hachenberger:2018:DMS}, generates exact results, but is slow for large models. A variant of this algorithm has been introduced in~\cite{Campen:2010:PBE}, where the arbitrary precision arithmetic is replaced by standard floating-point arithmetic. However, the performance of this method is still insufficient to achieve interactive runtimes.
Offset surfaces can be used for shape optimization, e.g., to optimize weight distribution inside an object~\cite{Musialski:2015:ROS}.

\paragraph{Resampled Offsets}
To avoid the explicit computation of the Minkoswki sum, which is a challenging and time-consuming operation, other methods resample the offset surface by computing the isosurface of the signed-distance function of the original surface~\cite{Varadhan:2006:AMS,Peternell:2007:MSB}.
Other methods relying on adaptive sampling include~\cite{Pavic:2008:HRV,Liu:2011:FIF}, but they are known to have a large memory overhead when a high accuracy is required.
\textcite{Meng:2017:ECF} distributes and optimizes the position of sites at a specified distance from the original surface, and produces a triangle from a restricted Delaunay triangulation of the sites. \textcite{Calderon:2014:PMO} introduced a framework for performing morphology operations directly on point sets.

\paragraph{Voronoi Diagrams and Signed-Distance Transforms}
Centroidal tessellations of Voronoi and power diagrams have important application in geometry processing and remeshing~\cite{Liu:2009:OCV,Xin:2016:CPD}. Our algorithm relies on these techniques, in particular on the implicit computation of Voronoi and power diagrams of points and segments.

\textcite{Fortune:1987:ASA} introduces a sweep line algorithm to position the Voronoi vertices (intersection of 3 bisectors) and create a 2D Voronoi diagram of point sets.
By relying on a slight modification of the Voronoi diagram definition, called half-space Voronoi diagrams~\cite{Fan:2011:HPV}, we will see in \Cref{sec:method} how to compute the Voronoi diagram of a set of parallel segments directly, using two sweeps instead of one, and how it leads to the efficient computation of a discrete offset surface. The extension of our method to 3D requires the computation of power diagrams~\cite{Aurenhammer:1987:PDP} with a weight associated to each seed.
Our algorithm implicitly relies on the Voronoi and power diagrams of line segments: while those could be approximated by sampling each segments with multiple points~\cite{Lu:2012:CVT}, we opted for an alternative solution which is more efficient and simpler to implement.

Finally, our 3 dimensional, two-stage sweeping algorithm bears some similarity to existing signed-distance transform approaches~\cite{Meijster:2002:AGA,Maurer:2003:ALT}, as it leverages the separability of the Euclidean distance to compute the resulting offset by sweeping in two orthogonal directions. However, as we operate on a dexel structure, we never need to store a full 3D distance field in memory. For a more complete review of distance transform algorithms, the reader is referred to the survey~\cite{Fabbri:2008:DED}.

\begin{figure}[tbp!]
	\centering

	\makebox[\linewidth][c]{
	\subfloat[Input Surface.]{
		\includegraphics[width=0.33\linewidth]{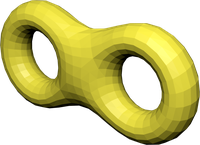}
		\label{fig:dexels:input}
	}
	\subfloat[Dexel Approximation.]{
		\includegraphics[width=0.33\linewidth]{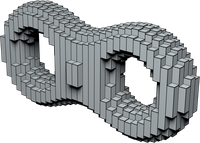}
		\label{fig:dexels:output}
	}
	\subfloat[Compact Storage.]{
		\includegraphics[width=0.33\linewidth]{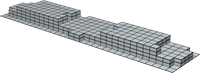}
		\label{fig:dexels:compact}
	}
	}

	\caption{
		A dexel data structure constructed from a triangle mesh. The input surface \protect\subref{fig:dexels:input} intersects with evenly spaced parallel rays \protect\subref{fig:dexels:output}, and is stored compactly as a 2D grid of events in the dexel buffer \protect\subref{fig:dexels:compact}.
		This can be used to perform efficient CSG operations in modeling software \cite{Lefebvre:2013:IAG}.
	}
	\label{fig:dexels}
\end{figure}

\rev{
\paragraph{Medial Axis and Skeleton}
Medial axis and shape skeletons are widely used in geometry processing applications, such as shape deformation, analysis, and classification~\cite{Tagliasacchi:2016:DSA}. One popular approach is \enquote{thinning}~\cite{Lam:1992:TMA,Liu:2010:ASA}, which exploits an erosion operator to reduce a shape to its skeleton. Our algorithm provides a practical and efficient way of defining such an operator. A shape can be approximated by a union of balls centered on its medial axis, and Voronoi diagrams are a common way of computing candidate centers for these balls~\cite{Amenta:2001:TPC}. These algorithms are know to be sensitive to small scale surface details~\cite{Attali:2009:SAC}, and special care needs to taken to ensure robustness~\cite{Yan:2016:ETO,Yan:2018:VCO}.
}

\paragraph{Ray-Based Representations and Offsets}
\label{par:relwork:fabrication:dexels}
A ray-based representation of a shape is obtained by computing the intersection of a set of rays with the given shape. In most applications, the cast rays are parallel and sampled on a uniform 2D grid. They are typically stored in a structure called \emph{dexel buffer}. A dexel buffer is simply a 2D array, where each cell contains a list of \emph{intersection} events $(\zIn, \zOut)$, each one representing one intersection with the solid (\Cref{fig:dexels}).
To the best of our knowledge, the term dexel (for \emph{depth pixel}) can be traced back to~\cite{VanHook:1986:RTS}, which introduced the dexel buffer to compute the results of CSG operations to ease NC milling path-planning.
Similar data structures have been described in different contexts over the years.
\emph{Layered Depth Images} (LDI)~\cite{Shade:1998:LDI} are used to achieve efficient image-based rendering.
The A-buffer~\cite{Carpenter:1984:TAB,Maule:2011:ASO} was used to achieve order-independent transparency.
It is worth mentioning that, while the construction algorithm is different, the underlying data structures used in all these algorithms remain extremely similar.
The G-buffer~\cite{Saito:1991:NMW} stores a normal in every pixel for further image-processing and CNC milling applications, augmenting ray-reps with normal information. In the context of additive manufacturing, \emph{Layered Depth Normal Images} have been proposed as an alternative way to discretize 3D models~\cite{Wang:2010:SMO,Huang:2014:AFL}.

Ray-based data structures offer an intermediate representation between usual boundary representations (such as triangle meshes), and volumetric representations (such as a full or sparse 3D voxel grids).
While both 3D images and dexel buffers suffer from uniform discretization errors across the volume, a dexel buffer is cheaper to store compared to a full volumetric representation: it is possible to represent massive volumes ($2048^3$ and more) on standard workstations.
Additionally, in the context of digital fabrication, 3D printers and CNC milling machines have limited precision: a dexel buffer at the resolution of the printer is sufficient to cover the space of shapes that can be fabricated.

Ray-reps have been used to compute and store the results of Minkoswki sums, offsets, and CSG operations~\cite{Hartquist:1999:ACS}.
\textcite{Hui:1994:SSI} uses ray-reps to compute a solid sweeping in image-space.
In~\cite{Chen:2011:UOO},~\citeauthor{Chen:2011:UOO} use LDNI to offset polygonal meshes by filtering the result of an initial overestimate of the true dilated shape.
\textcite{Wang:2013:GBO} computes the offset mesh as the union of spheres placed on the points sampled by the LDI\@.
Finally,~\cite{Martinez:2015:CSO} approximates the dilation by a spherical kernel with a zonotope, effectively computing the Minkoswki sum of the original shape with a sequence of segments in different directions. Differently, the method presented in this paper computes the \emph{exact offset} of the discrete input ray-rep (at the resolution of the dexel representation).
Despite using a similar data structure, our approach is different from \textcite{Wang:2013:GBO}: the latter requires a LDI sampled from 3 orthogonal directions, while our method accelerates the offset operation even when a ray-rep from a single direction is available. We provide a comparison and describe the differences in more details in \Cref{sec:results}.

\begin{figure}[tbp!]
	\centering

	\includegraphics[width=0.6\linewidth]{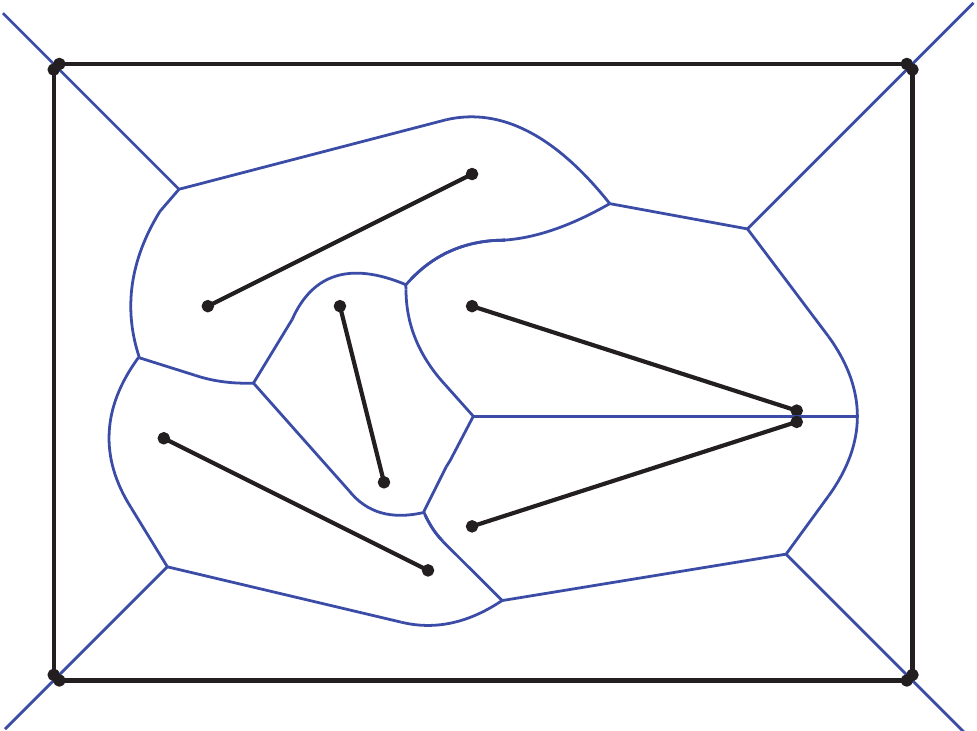}

	\caption{
		Voronoi diagram of line segments. The bisectors are defined by second-order piecewise polynomial curves.
		Note that the Voronoi cell of a single seed can be comprised of multiple connected components.
		Illustration from \cite{vanKreveld:2017:LNI}.
	}
	\label{fig:voronoi-segments}
\end{figure}

\paragraph{Voronoi Diagrams of Line Segments and Motivations}
An example of a Voronoi diagram of line segments is shown in \Cref{fig:voronoi-segments}. Although the general abstract algorithm for Voronoi diagrams applies (in theory) to line segments~\cite{Fortune:1987:ASA}, the problem is in practice extremely challenging, since the bisector of two segments are piecewise second-order polynomial curves. Some software is readily available to solve the 2D case (e.g., \rev{VRONI~\cite{Held:2001:VAE,Held:2009:TOI}, Boost.Polygon~\cite{Boost:2010:TBP}, OpenVoronoi~\cite{Wallin:2018:OPE}} and CGAL~\cite{Karavelas:2018:DSD}), but the problem is still open in 3D.
\rev{\textcite{Aurenhammer:2017:VDF} present a method for computing the \emph{Voronoi diagram} of \emph{parallel} line segments by reducing the problem to computing the \emph{power diagram} of points in a hyperplane, but the problem of computing the \emph{power diagram of line segments} is still open, even in 2D.}
To the best of the authors' knowledge, the problem of computing the \rev{\emph{power diagram for a set of line segments}} was not explored in the literature, not even in 2D.
This component is necessary for the second step of our dilation process, and it is one of the contributions of this work. Note that our method does not explicitly store a full Voronoi or power diagram at any time, since the process is \emph{online} (the cells are computed for the current sweepline only and updated as our sweep progresses).

Compared to existing offsetting techniques, there are several advantages to using our discrete approach when the result needs only be computed at a fixed resolution (e.g., the printer resolution in additive manufacturing).

\noindent\textbf{(1)} No need for a clean input triangle mesh. As long as the input triangle soup can be properly discretized (e.g., using generalized winding number~\cite{Jacobson:2013:RIO,Barill:2018:FWN}), the input can have gaps or self-intersecting triangles. %

\noindent\textbf{(2)} Our method is also directly applicable when the input is a 3D image (e.g., CT scan), where it can be used without any loss of accuracy, providing high performance and low memory footprint.

\noindent\textbf{(3)} Another advantage of working directly with the dexel data structure is that CSG operations can be carried out easily in dexel space, providing real-time interactive modeling capabilities~\cite{Lefebvre:2013:IAG}. Performing CSG operations on triangle meshes is costly and requires a significant implementation effort to be performed robustly~\cite{Zhou:2016:MAF}. Compared to~\cite{Wang:2013:GBO}, our method can achieve higher resolutions when needed, and compared to~\cite{Martinez:2015:CSO}, the result of the dilation by a spherical kernel is computed exactly.

\section{Method}
\label{sec:method}%

\begin{figure}[tbp!]
	\centering

	\hfill{}
	\subfloat[]{\begin{minipage}[c][4cm]{3cm}
		\centering
		\includegraphics[height=4cm]{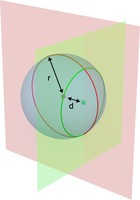}
		\label{fig:twopass:1}
	\end{minipage}}
	\hfill{}
	\subfloat[$\shapeIn$]{\begin{minipage}[c][4cm]{1cm}
		\centering
		\includegraphics[height=.4cm]{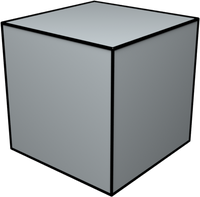}
		\label{fig:twopass:2}
	\end{minipage}}
	\hfill{}
	\subfloat[$\shapeMid$]{\begin{minipage}[c][4cm]{2cm}
		\centering
		\includegraphics[height=0.8cm]{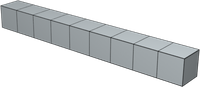}
		\label{fig:twopass:3}
	\end{minipage}}
	\hfill{}
	\subfloat[$\shapeOut$]{\begin{minipage}[c][4cm]{1.8cm}
		\centering
		\includegraphics[height=1.8cm]{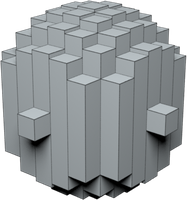}
		\label{fig:twopass:4}
	\end{minipage}}
	\hfill{}

	\caption{
		The dilation operation is performed in two stages \protect\subref{fig:twopass:1}.
		A seed is dilated by a radius $\dilationRadius{}$ along a first axis (in red).
		The resulting seed (in green) is then dilated along a second axis by a different radius $\sqrt{d^2 - \dilationRadius^2}$, where $d$ is the distance between the two seeds (red point and green point).
		The equivalent dexel data structure for each pass is shown on right: \protect\subref{fig:twopass:2} input dexel, \protect\subref{fig:twopass:3} result of the first pass, \protect\subref{fig:twopass:4} result of the second pass.
	}
	\label{fig:twopass-detail}
\end{figure}

The key idea of our algorithm is to decompose the dilation of a point in two 2D dilations along orthogonal axes. \Cref{fig:twopass:1} illustrates this idea. The result of the 3D dilation of a point---the blue sphere \rev{in \Cref{fig:twopass:1}}---can be computed by first performing the dilation along the first axis (in red). Then, the green point can be dilated along a second axis (in green) by a different radius, producing the green circle. By applying this construction directly to a dexel buffer, the input point in \Cref{fig:twopass:2} is first dilated to produce the line \rev{in} \Cref{fig:twopass:3}. Then, each element of this line is dilated again in the orthogonal direction, but with a different dilation radius, producing the final result (\Cref{fig:twopass:4}). Each stage is a set of 2D dilations: the first stage relying on Voronoi diagrams (uniform offsetting) and the second stage leveraging power diagrams to compute efficient offsets against different dilation radii for each dexel segment. We denote by $\shapeIn \to \shapeMid \to \shapeOut$ the different steps of the pipeline.

We start by formally defining the different concepts and notations in \Cref{subsec:method:overview}.
In \Cref{subsec:method:voronoi-2d}, we describe our uniform offsetting method (in 2D), based on efficiently computing the Voronoi diagram of a set of parallel segments.
Then, in \Cref{subsec:method:power-3d}, we extend the algorithm to compute the power diagram of a set of parallel segments in 2D, which completes the necessary building blocks to extend our algorithm from 2D to 3D.

\subsection{Definitions}
\label{subsec:method:overview}

\paragraph{Dexel Representation}
A dexel buffer of a shape $\shape \subset \domain$ (with $\domain \eqdef \reals^3$) is a set of parallel segments arranged in a 2D grid, where each cell $\shape_{ij}$ of the grid contains a list of segments sharing the same $xy$ coordinate.
Specifically, we discretize $\shape$ as $\shape \approx \cup_{ij} \shape_{ij}$ where $\shape_{ij} = \{ (\zIn_k, \zOut_k) \}_{k=1}^{n_{ij}}$.
Each segment $(\zIn_k, \zOut_k)$ in the same cell $\shape_{ij}$ has the same $xy$ coordinate, and represents the intersections of the input shape $\shape$ with a vertical ray at the same $xy$ coordinate (see \Cref{fig:dexels}).

\paragraph{Dilation}
Given an input shape $\shape$ and a radius $\dilationRadius \geq 0$ we define the \emph{dilated shape} $\dilate_\dilationRadius{}(\shape)$ as the set of points at a distance less or equal than $\dilationRadius{}$ from $\shape$:
\begin{equation}
	\dilate_\dilationRadius{}(\shape) = \{ \point \in \domain, \norm{\point - \otherPoint} \leq \dilationRadius{}, \otherPoint \in \shape \}.
\end{equation}

\paragraph{Erosion}
The erosion of a shape $\shape$ is equivalent to computing the dilation on the complemented shape $\compl{\shape}$, and taking the complement of the result. Since the complement operation is trivial under a ray-rep representation, we will focus on describing our algorithm applied to the dilation operation, from which all other morphological operators (erosion, closing, and opening) will follow.

\paragraph{Voronoi Diagram}
The \emph{Voronoi diagram} of a set of \emph{seeds} $\seeds = \{\seed_i\}_{i=1}^n$ is a partition of the space $\domain$ into different \emph{Voronoi cells} $\voronoiCell_i$:
\begin{equation}
	\voronoiCell_i = \left\{ \point \in \domain, \dist(\point, \seed_i) \leq \dist(\point, \seed_j), i \neq j \right\}.
	\label{eq:voronoi-cell}
\end{equation}

We also define $\Vor{\seeds}$ to be the interface between each overlapping Voronoi cell:
\begin{equation}
	\Vor{\seeds} = \cup_{i \neq j} \voronoiCell_i \cap \voronoiCell_j.
\end{equation}

\emph{Notations.} In the context of this paper, seeds can be either points or line segments.
Throughout the paper, $\seed_i$ shall denote the geometric entity of a seed, whether it is a point or a line segment. $\seedPoint_i$ shall denote the position of the seed when it is a point, and $(\seedIn_i, \seedOut_i)$ shall be used to denote the positions of its endpoints when the seed is a line segment.
When the seeds \rev{$\{\seed_i\}_i$} are points, $\Vor{\seeds}$ is a set of straight lines in 2D (planes in 3D), which are equidistant to their closest seeds. When the seeds are segments, $\Vor{\seeds}$ is comprised of parabolic arcs in 2D~\cite{Lu:2012:CVT}, and parabolic surfaces in 3D.

\begin{figure}[tbp!]
	\centering

	\makebox[\linewidth][c]{
	\includegraphics[width=0.33\linewidth]{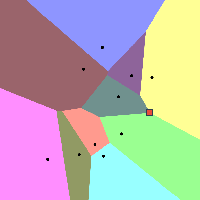}
	\begin{overpic}[width=0.33\linewidth]{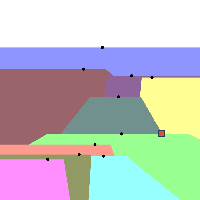}
		\put (10,88) {$\downarrow$}
	\end{overpic}
	\begin{overpic}[width=0.33\linewidth]{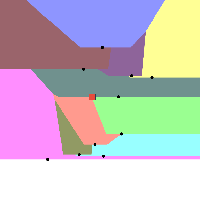}
		\put (80,5) {$\uparrow$}
	\end{overpic}
	}

	\caption{
		Voronoi diagram of seed points $\{\seed_i\}_{i=1}^n$.
		From left to right: Voronoi diagram formed by the full Voronoi cells $\voronoiCell_i$; half-space Voronoi diagram formed by the half-space Voronoi cells $\protect\Fwd{\voronoiCell}_i$ and $\protect\Rev{\voronoiCell}_i$ respectively.
		Note that $\voronoiCell_i \subseteq \protect\Fwd{\voronoiCell}_i \cup \protect\Rev{\voronoiCell}_i$.
		In each diagram, a Voronoi vertex (intersection between $3^+$ Voronoi cells) is shown with a red square.
	}
	\label{fig:half-voronoi}
\end{figure}

\begin{figure}[tbp!]
	\centering

	\makebox[\linewidth][c]{
	\includegraphics[width=0.33\linewidth]{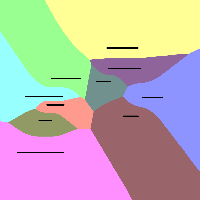}
	\begin{overpic}[width=0.33\linewidth]{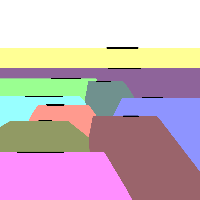}
		\put (10,88) {$\downarrow$}
	\end{overpic}
	\begin{overpic}[width=0.33\linewidth]{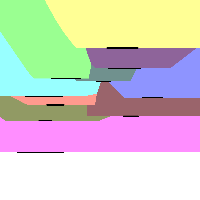}
		\put (80,5) {$\uparrow$}
	\end{overpic}
	}

	\caption{
		\rev{
		Voronoi diagram of seed segments $\{\seed_i\}_{i=1}^n$.
		From left to right: Voronoi diagram formed by the full Voronoi cells $\voronoiCell_i$; half-space Voronoi diagram formed by the half-space Voronoi cells $\protect\Fwd{\voronoiCell}_i$ and $\protect\Rev{\voronoiCell}_i$ respectively.
		}
	}
	\label{fig:half-voroseg}
\end{figure}

In a half-space Voronoi diagram for seed points~\cite{Fan:2011:HPV}, each seed point $\seed_i$ is associated with a \emph{visibility direction} $\visdir_i$, and a point $\point$ is considered in \Cref{eq:voronoi-cell} if and only if $(\point - \seedPoint_i) \cdot \visdir_i \geq 0$.
In this work, we are interested in half-space Voronoi diagrams of seed segments, where each seed is associated to the same visibility direction $\visdir$.
\Cref{fig:half-voronoi} \rev{(resp.\ \Cref{fig:half-voroseg})} shows the difference between Voronoi diagrams and half-space Voronoi diagrams for point seeds \rev{(resp.\ segment seeds)}.
More precisely, given a set of parallel seed segments $(\seedIn_i, \seedOut_i)_{i=1}^n$, and a direction $\visdir$ orthogonal to each segment $\visdir \perp (\seedOut_i - \seedIn_i)$, we define the half-space Voronoi cells $\Fwd{\voronoiCell}_i$ and $\Rev{\voronoiCell}_i$ as
\begin{equation}
\begin{aligned}
	\Fwd{\voronoiCell}_i = \big\{ &\point \in \domain, \norm{\point - \proj{\point}_i} \leq \norm{\point - \proj{\point}_j}, i \neq j, \\
			& (\point - \proj{\point}_i) \cdot \visdir \geq 0,
			(\point - \proj{\point}_j) \cdot \visdir \geq 0 \big\} \\
	\Rev{\voronoiCell}_i = \big\{ &\point \in \domain, \norm{\point - \proj{\point}_i} \leq \norm{\point - \proj{\point}_j}, i \neq j, \\
			& (\point - \proj{\point}_i) \cdot \visdir \leq 0,
			(\point - \proj{\point}_j) \cdot \visdir \leq 0 \big\}
\end{aligned}
\label{eq:half-space-voronoi-cell}
\end{equation}
where $\proj{\point}_i$ is the point $\point$ projected on the line $(\seedIn_i, \seedOut_i)$.
Note that the segments $(\seedIn_i, \seedOut_i)$ are parallel, \rev{following a direction that is orthogonal to} the chosen direction $\visdir$. \rev{Thus,} the dot product $(\point - \vec{q}) \cdot \visdir$ in \Cref{eq:half-space-voronoi-cell} has the same sign for all the points $\vec{q}$ in the line $(\seedIn_i, \seedOut_i)$.
Note that we have ${\voronoiCell_i \subseteq \Fwd{\voronoiCell_i} \cup \Rev{\voronoiCell_i}}$ (see \Cref{fig:half-voronoi}).

\paragraph{Half-Dilated Shape}
We define the \emph{half-dilated shape} $\Fwd{\dilate}(\shape)$, as the dilation restricted to the half-space Voronoi cells of segments in $\shape$:
\begin{equation}
	\Fwd{\dilate}_\dilationRadius{}(\shape) = \{ \point \in \Fwd{\voronoiCell}_i, \norm{\point - \otherPoint} \leq \dilationRadius{}, \otherPoint \in \seed_i, i \in \irange{1, n} \}.
\end{equation}
Remember that $\seed_i$ is the $i$-th segment $(\zIn_i, \zOut_i)$ in the dexel buffer approximating the shape $\shape$. The half-space dilated shape $\Rev{\dilate}_\dilationRadius{}(\shape)$ is defined in a similar manner using $\Rev{\voronoiCell}_i$.
For simplicity, we will omit the dilation radius and simply write $\dilate(\shape)$, unless there is an ambiguity.

\paragraph{Power Diagram}
Finally, the \emph{power diagram} of a set of seeds $\{\seed_i\}_{i=1}^{n}$ is a weighted variant of the Voronoi diagram. Each seed is given a weight $\weight_i$ that determines the size of the \emph{power cell} $\powerCell_i$ associated to it:
\begin{equation}
	\powerCell_i = \left\{ \point \in \domain, \dist(\point, \seed_i)^2 - \weight_i \leq \dist(\point, \seed_j)^2 - \weight_j, i \neq j \right\}.
	\label{eq:power-region}
\end{equation}
Intuitively, one could interpret a 2D power diagram as the orthographic projection of the intersection of a set of parabola centered on each seed, where the weights determine the height of each seed embedded in $\reals^3$. This definition extends naturally to \emph{half-space power diagrams}.

\subsection{Half-Space Voronoi Diagram of Segments and 2D Offsets}
\label{subsec:method:voronoi-2d}

Given a 2D input shape $\shape$ and dilation radius $\dilationRadius{}$, we seek to compute the dilated shape $\dilate(\shape)$ comprised of the set of points at a distance $\leq \dilationRadius{}$ from $\shape$. The input shape is given as a union of disjoint parallel segments evenly spaced on a regular grid (the dexel data structure), and we seek to compute the output shape as another dexel structure %
(the discretized version of the continuous dilated shape). %

The dilated shape $\dilate(\shape)$ can be expressed as the union of the dilation of each individual segment of $\shape$. To compute this union efficiently, our key insight is to partition the dilated shape based on the Voronoi diagram of the input segments \rev{(power diagrams are only needed for the extension to 3D)}. Within each Voronoi cell $\voronoiCell_i$, if a point $\point$ is at a distance $\leq \dilationRadius{}$ from the seed segment $(\seedIn_i, \seedOut_i) \in \shape$, then $\point \in \dilate(\shape)$.

\begin{figure}[tb]
	\centering

	\hfill{}
	\includegraphics[width=0.4\linewidth]{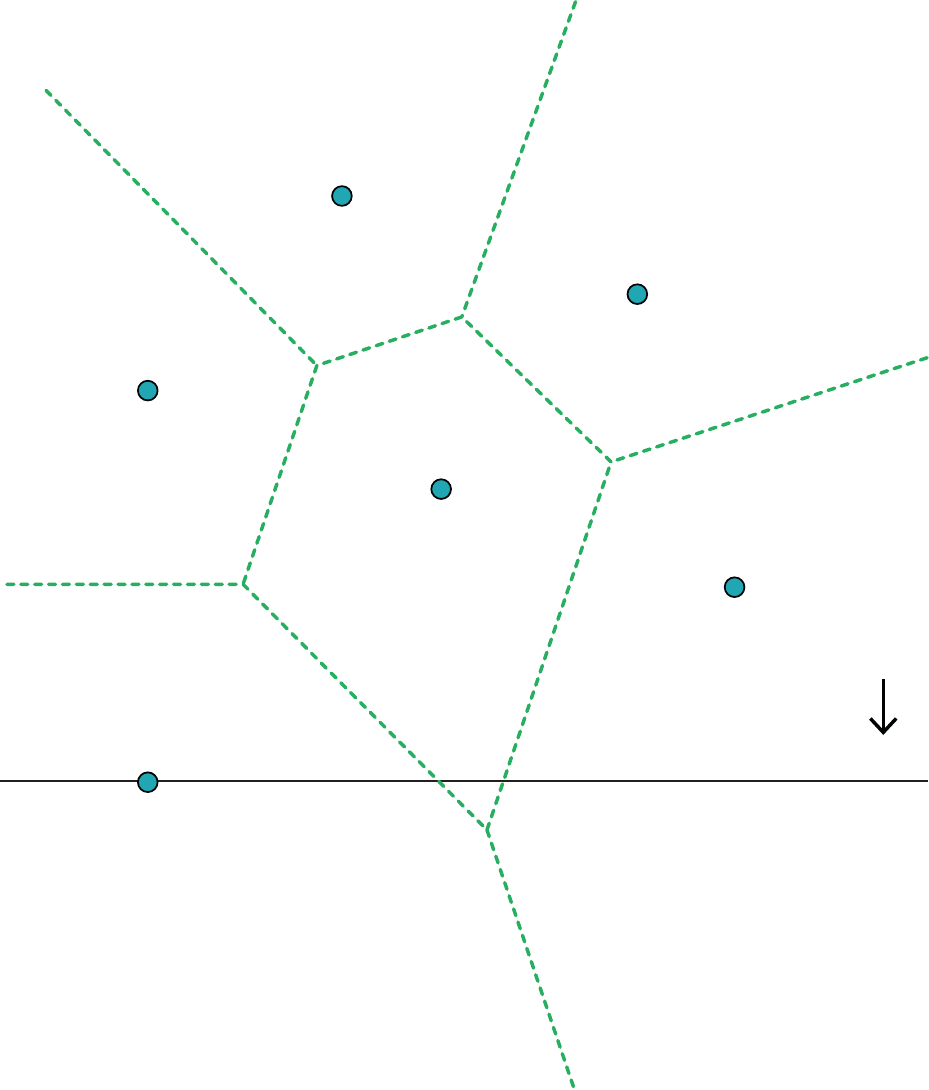}\hfill{}
	\includegraphics[width=0.4\linewidth]{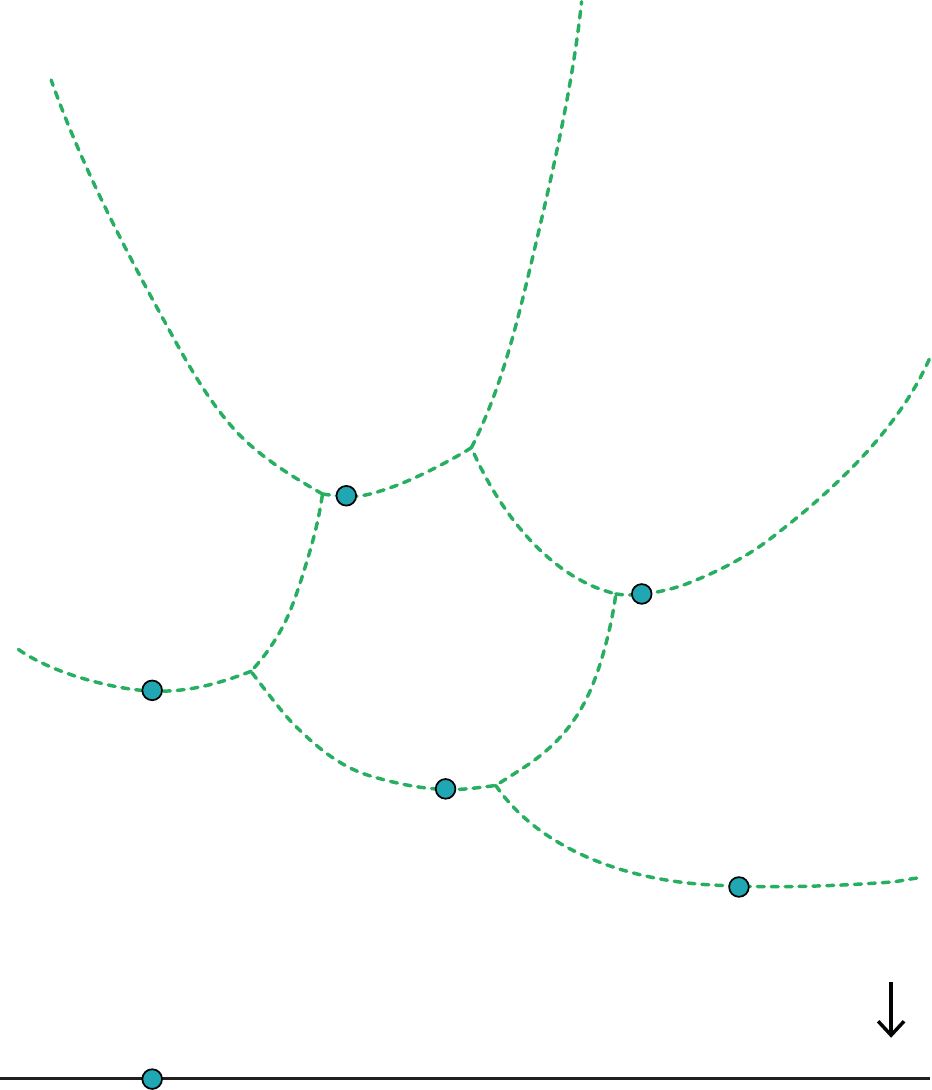}\hfill{}

	\caption{
		Fortune's sweepline algorithm \cite{Fortune:1987:ASA} requires transforming the point coordinates to compute the correct Voronoi diagram of points in a single sweeping pass, which makes it impossible to compute the result of the dilation in the same sweeping pass \rev{(without back-propagating the contribution of each newly inserted seed in the dilated shape)}.
	}
	\label{fig:fortune}
\end{figure}

The Voronoi diagram of point seeds can be computed in single pass with a sweepline algorithm~\cite{Fortune:1987:ASA}. The construction requires lifting coordinates in the plane according to the coordinate along the sweep direction, as illustrated in \Cref{fig:fortune}.
Unfortunately, this approach cannot be used to compute the result of a dilation operation in a single pass \rev{without \enquote{backtracking}. Indeed, whenever a new seed is added to the current sweepline, its dilation will affect rows of the image above the sweepline, which have already processed.}
Instead, we propose a simple construction for seed points and segments, based on half-space Voronoi diagrams---which we extend to power diagrams in \Cref{subsec:method:power-3d}.
Our key idea is to compute the dilation of a segment $(\seedIn_i, \seedOut_i)$ in its Voronoi cell $\voronoiCell{}_i$ as the union of the two half-dilated segments, in $\Fwd{\voronoiCell{}}_i$ and $\Rev{\voronoiCell{}}_i$. Both $\Fwd{\voronoiCell{}}_i$ and $\Rev{\voronoiCell{}}_i$ can be computed efficiently in two separate sweeps of opposite direction, without requiring any transformation to the coordinate system as in~\cite{Fortune:1987:ASA}.

\begin{figure}[tbp!]
	\centering

	\tcenter{\includegraphics[width=0.6\linewidth]{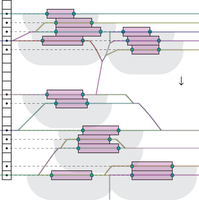}}
	~
	\tcenter{\frame{\includegraphics[width=0.2\linewidth]{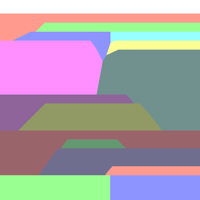}}}

	\caption{
		A single sweep in one direction allows us to compute the half-space Voronoi diagram of parallel line segments (the dexel data structure).
		This figure illustrates how dexels are stored as nested arrays. The gray area shows the expected result of the half-dilation in one direction, \rev{while the solid colored lines show the boundary of the half-space Voronoi diagram of the input line segments (half-space Voronoi cells are shown on the top right)}.
	}
	\label{fig:morpho:sweep}
\end{figure}

\begin{figure}[tbp!]
	\centering

	\begin{overpic}[width=0.8\linewidth]{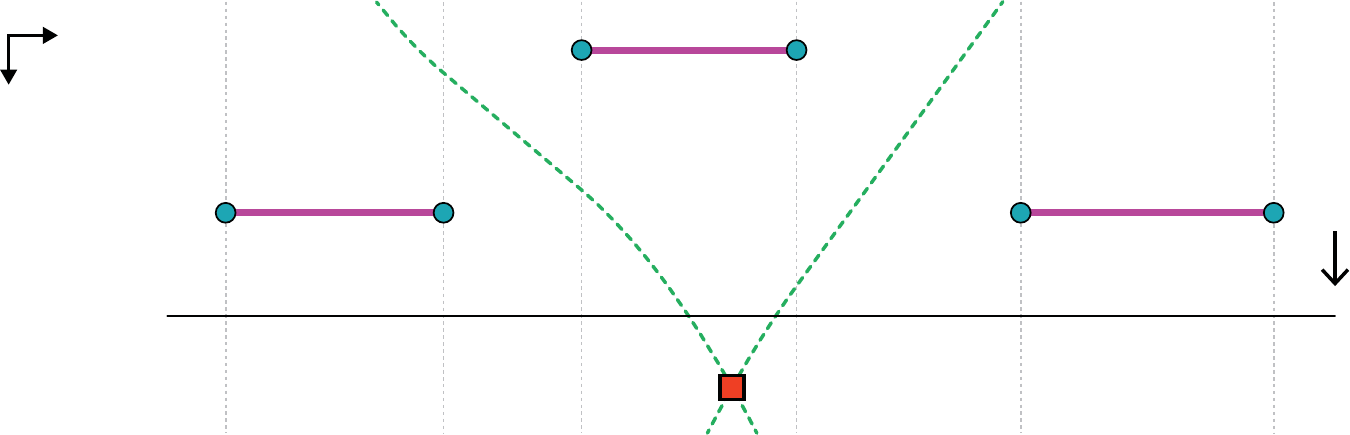}
		\put (-1,22) {$y$}
		\put (6,29) {$z$}
	\end{overpic}

	\caption{
		The bisector of two line segments can be described by a piecewise second-order polynomial curve.
		Voronoi vertices are located at the intersection between two such bisectors.
		After this point, the middle segment will always be \emph{further} away from the sweepline than its two neighbors.
		It will be marked as \emph{inactive} and be removed from $\mL$.
	}
	\label{fig:voronoi-vertex}
\end{figure}

The idea behind our sweeping algorithm is as follows. We advance a sweepline $L$ parallel to the seed segments in the input dexel $\shape$. A general view of a 2D dexel buffer and the sweep process is presented in \Cref{fig:morpho:sweep}, while \Cref{fig:voronoi-vertex} shows the structure of the Voronoi diagram of three line segments during the sweep process.
At each step, for each seed \rev{$\seed_i \in \shape$}, we compute the intersection of the current line $L$ with the points in $\point \in \Fwd{\voronoiCell{}}_i$ that are at a distance $\leq \dilationRadius{}$. By choosing the visibility direction $\visdir$ to be the same as the sweeping direction, then only the seeds \rev{$\{\seed_i\}_i$} previously encountered in the sweep will contribute to this intersection (\rev{any} upcoming seed will have an empty Voronoi cell $\Fwd{\voronoiCell{}}_i$).
\Cref{fig:half-voronoi} shows a Voronoi diagram of points, with two half-space Voronoi diagrams of opposite directions.
To make the computation efficient, we do not want to iterate through all the seeds to compute the intersection each time we advance the sweep line (which would make the algorithm $\O(n^2)$ in the number of seeds). Instead, we want to keep only a small list of \emph{active seeds}, that will contribute to the dilated shape $\dilate(\shape)$ and intersect the current sweepline $L$: this will decrease the complexity to $\O(n \log(n) + m)$, where $m$ is the number of line segments generated by the dilation process.

\paragraph{Pseudocode}
From an algorithmic point of view, we maintain, during the sweeping algorithm, two data structures: $\mL$, the list of \emph{active} seed segments $(\seedIn_i, \seedOut_i)$, whose Voronoi cell $\Fwd{\voronoiCell{}}_i$ intersects the current sweepline, (and are thus at distance $\leq \dilationRadius{}$ from the current sweepline), and $\mQ$, a priority queue of upcoming \emph{events}. These events indicate when an active seed can safely be removed from the $\mL$ as we advance the sweepline, i.e., when it no longer affects the result of the dilation.
The events in $\mQ$ can be of two types: (1) a Voronoi vertex (the intersection between $3^+$ Voronoi cells); and (2) a seed becoming inactive due to a distance $> \dilationRadius{}$ from the sweepline.
Voronoi vertices can be located at the intersection of the bisector curves between 3 consecutive seed segments $a, b, c \in \mL$, as illustrated in \Cref{fig:voronoi-vertex}. Beyond this intersection, we know that either $a$ or $c$ will be closer to the sweepline, so we can remove $b$ from $\mL$ (its Voronoi cell $\Fwd{\voronoiCell{}}_b$ will no longer intersect the sweepline).

Both $\mQ$ and $\mL$ can be represented by standard STL data structures. Note that since the seeds stored in $\mL$ are \emph{disjoints}, they can be stored efficiently in a \inlinecode{std::set<>} (sorted by the coordinate of their midpoint).
A detailed description of our sweep-line algorithm is given in pseudocode in \Cref{alg:voronoi:sweep-line,alg:voronoi:insert-segment}.
\rev{A full C++ implementation is available on \fnurl{https://github.com/geometryprocessing/voroffset}{github}.}
In \lref{alg:voronoi:sweep-line:dilate-line} \rev{in \Cref{alg:voronoi:sweep-line}}, the function \function{DilateLine} computes the result of the dilation on the current sweep line, by going through the list of active seeds, and merging the resulting dilated \emph{line} segments.
Insertion and removal of seed segments in $\mL$ is handled by the functions \function{InsertSeedSegment} and \function{RemoveSeedSegment} respectively.
When inserting a new active \emph{seed} segment in $\mL$, to maintain the efficient storage with the \inlinecode{std::set<>}, we need to remove subsegments which are occluded by the newly inserted seed segment (and split partially occluded segments). Indeed, the contribution of such seed segments is superseded by the new seed segment that is inserted. Then, we need to compute the possible Voronoi vertices formed by the newly inserted seed segment and its neighboring seeds in $\mL$.
The derivation for the coordinates of the Voronoi vertex of 3 parallel seed segments in given in \Cref{app:voronoi-vertex}.

\begin{figure}[htbp!]
	\begin{algorithmic}[1]
		\Require
		2D dexel structure $\shape$ + dilation radius $\dilationRadius{}$.

		\Ensure
		Half-dilated dexel shape $\otherShape{} = \Fwd{\dilate}_\dilationRadius(\shape)$.

		\Function{VoronoiSweepLine}{$\shape, \dilationRadius{}$}
		\State $\mL \gets \emptyset$ \Comment{Set of active segments on the sweep line}
		\State $\mQ \gets \{ \}$ \Comment{List of removal events marking a seed as \emph{inactive}}
		\State $\otherShape \gets \emptyset$ \Comment{Dilated result}
		\For{$\mathtt i \gets 0, N-1$}
		\ForAll{$\seed_j \in \mQ$}
		\State \function{RemoveSeedSegment}$(\mL, \dilationRadius, \seed_j)$; \label{alg:voronoi:sweep-line:remove-seed}
		\EndFor
		\ForAll{$\seed_j^i = (\z\sIn, \z\sOut) \in \shape_i$}
		\State \function{InsertSeedSegment}$(\mL, \mQ, \dilationRadius, \seed_j)$; \label{alg:voronoi:sweep-line:insert-seed}
		\EndFor
		\State \Comment{Dilate and merge active seeds on the current sweepline}
		\State $\otherShape \gets \otherShape \cup$ \function{DilateLine}$(\mL, i, \dilationRadius)$ \label{alg:voronoi:sweep-line:dilate-line}
		\EndFor
		\State \Return{$\otherShape$}
		\EndFunction
	\end{algorithmic}

	\caption{Sweepline algorithm for computing the half-dilated shape $\protect\Fwd{\dilate}_\dilationRadius{}(\shape)$.}
	\label{alg:voronoi:sweep-line}
\end{figure}

\begin{figure}[htbp!]
	\begin{algorithmic}[1]
		\Require
		\begin{tabular}{@{\hspace{-0.25em}}l@{\hspace{-1em}}l@{~}l}
		    \multirow{4}{*}{$\begin{cases} \\ \\  \\ \end{cases}$}
			& $\mL$ & Set of active seeds on the sweep line, \\
			& $\mQ$ & List of removal events, \\
			& $\dilationRadius$ & Dilation radius, \\
			& $\seed$ & Seed segment to insert, $\seed = \segment{(y, z\sIn), (y, z\sOut)}.$
		\end{tabular}

		\Ensure
		Updated list of active seeds $\mL$ and events $\mQ$.

		\Function{InsertSeedSegment}{$\mL, \mQ, \dilationRadius, \seed$}
		\State \function{RemoveOccludedSegments}$(\mL, \seed)$
		\State \function{SplitPartiallyOccluded}$(\mL, \seed)$
		\State $\mL \gets \mL \cup \seed$ \Comment{Overlaps are resolved, insert $\seed$ into $\mL$}
		\State $\mQ \gets \mQ \cup (y + \dilationRadius, \seed)$ \Comment{After this point, $\dilate_\dilationRadius{}(\seed)$ will be empty}
		\While{$\exists$ sequence $(\seed_a, \seed_b, \seed) \in \mL$}
		\State $(y_v, z_v) \gets$ \function{VoronoiVertex}$(\seed_a, \seed_b, \seed)$ \Comment{See \cref{fig:voronoi-vertex}}
		\If{$y_v < y$}
		\State $\mL \gets \mL \setminus \seed_b$ \Comment{Seed $\seed_b$ becomes inactive}
		\Else
		\State$\mQ \gets \mQ \cup (y_v, \seed_b)$ \Comment{Remove $\seed_b$ later on}
		\EndIf
		\EndWhile
		\While{$\exists$ sequence $(s, s_a, s_b) \in \mL$}
		\State \Comment{Repeat operation on the right side of $s$}
		\EndWhile
		\EndFunction
	\end{algorithmic}

	\caption{Insertion of a new seed segment $\seed$ into $\mL$.}
	\label{alg:voronoi:insert-segment}
\end{figure}

\subsection{Half-Space Power Diagram of Points and 3D Offsets}
\label{subsec:method:power-3d}

\paragraph{3D Dilation}
As illustrated in \Cref{fig:twopass-detail}, the 3D dilation process is decomposed in two stages $\shapeIn{} \mapsto \shapeMid{} \mapsto \shapeOut{}$. We first perform an extrusion along the first axis (in red), followed by a dilation along the second axis (in green).
Note that the first operation $\shapeIn{} \mapsto \shapeMid{}$ is not exactly the same as a dilation along the first axis (red plane): the segments $(\seedIn_i, \seedOut_i) \in \shapeIn$ are \emph{extruded}, not dilated (they map to a rectangle, not a disk).

To obtain the final dilated shape in 3D, we need to perform a dilation of the intermediate shape $\shapeMid{}$, where  each segment $(\seedIn_j, \seedOut_j) \in \shapeMid$ is dilated by a different radius $\dilationRadius_j$ along the second axis (green plane in \Cref{fig:twopass-detail}), depending on its distance from parent seed (red dot in \Cref{fig:twopass-detail}).
In the case where the first extrusion of $\shapeIn{}$ produces overlapping segments in $\shapeMid{}$, the overlapping subsegments would need to be dilated by different radii, depending on which segment in $\shapeIn$ it originated from.
In such a case, where a subsegment $\subseed_j \in \shapeMid$ has multiple parents $\subseed_i \in \shapeIn$, it is enough to dilate $\subseed_j$ by the radius of its closest parent in $\shapeIn$ (the one for which $\dilationRadius_j^i = \sqrt{d_{ij}^2 - \dilationRadius^2}$ is the largest).
In practice, we store $\shapeMid$ as a set of non-overlapping segments, as computed by the algorithm in \Cref{alg:voronoi:sweep-line}, with a slight modification to the \function{DilateLine} function (\lref{alg:voronoi:sweep-line:dilate-line} \rev{in \Cref{alg:voronoi:sweep-line}}) to return the \emph{extruded} seeds on the current sweepline instead of the dilated ones.

\begin{figure}[tbp!]
	\centering
	\includegraphics[width=0.75\linewidth]{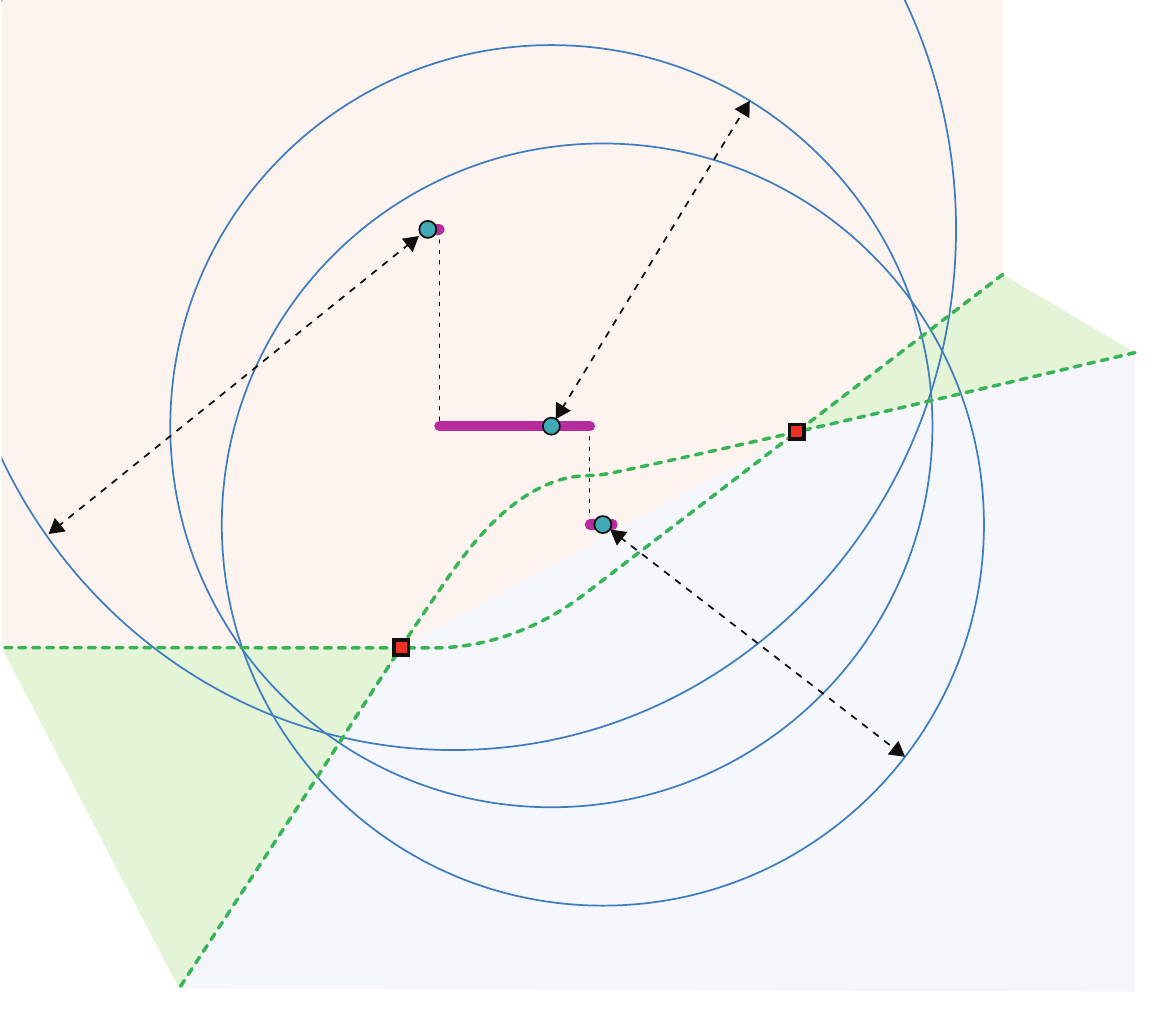}

	\caption{
		Special case: the power cell \rev{(in green)} of a seed segment \rev{(central segment in purple)} can be a disconnected region of the plane. \rev{The blue points represent the center of the power circles (the boundary of the power diagram is the locus of the lines intersecting each pair of circle as the centers move along their respective segment).}
	}
	\label{fig:power-segment}
\end{figure}

\paragraph{2D Power Diagram}
We now focus on computing the 2D dilation of a shape $\shape$ composed of non-overlapping segments $(\seedIn_i, \seedOut_i; \dilationRadius_i)$\rev{, where each seed segment is weighted by the dilation radius ($\weight_i \eqdef \dilationRadius_i$ in \Cref{eq:power-region})}.
In this setting, the computation of the power cells $\powerCell_i$ of seed segments becomes  more involved, breaking some of the assumptions of the sweepline algorithm.
Specifically, the sweepline algorithm (\Cref{alg:voronoi:sweep-line}) makes the following assumptions about $\mL$: the seeds projected on the sweepline are non-overlapping segments, and the Voronoi cells induced by the active seeds are \emph{continuous} regions.
Once a seed $(\seedIn_i, \seedOut_i)$ is inserted in $\mL$, its cell $\Fwd{\voronoiCell}_i$ immediately becomes \emph{active}, and once we reach the first removal event in $\mQ$, it will become \emph{inactive} and stop contributing to the dilation $\Fwd{\dilate}(\shape)$.
For the power cells of \emph{segments}, the situation is a little bit different, as illustrated on \Cref{fig:power-segment}: a power cell $\Fwd{\powerCell}_i$ \rev{of a line segment} can contain \emph{disjoints} regions of the plane.
It is not clear how to maintain a disjoint set of seeds in $\mL$ if we need to start removing and inserting a seed multiple time, and this makes the number of cases to consider grow significantly.

\begin{figure}[tbp!]
	\centering

	\begin{overpic}[width=0.8\linewidth]{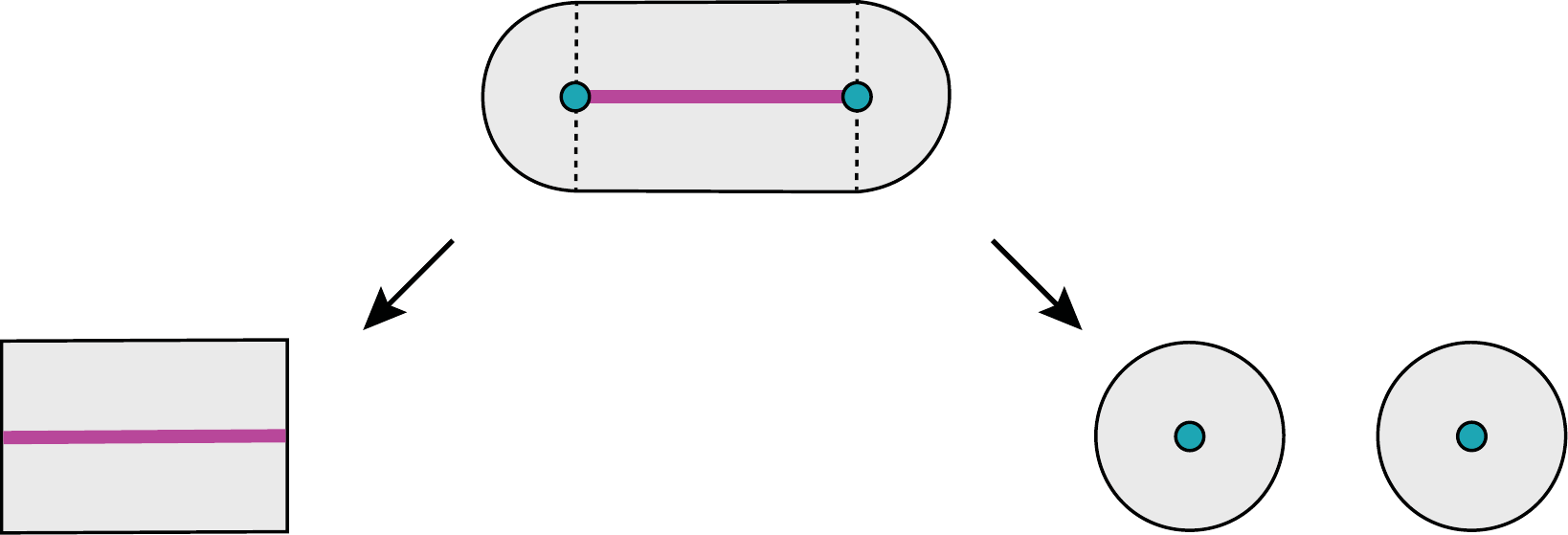}
		\put (15,30) {$\dilatePower({\shape})$}
		\put (3,15) {$\dilateSegments(\shape)$}
		\put (76,15) {$\dilateEndpoints(\shape)$}
	\end{overpic}

	\caption{
		To simplify the computation of the power diagram in the second stage, the dilation of a segment is separated into a vertical extrusion (left), and the dilation of its two endpoints (right).
	}
	\label{fig:segments-separate}
\end{figure}

Instead, we propose to circumvent the problem entirely by making the following observation.
A 2D segment dilated by a radius $\dilationRadius{}$ is actually a capsule, which can be described by two half-disks at the endpoints, and a rectangle in the middle.
We decompose the half-dilated shape $\Fwd{\dilatePower}(\shape)$ into the union of two different shapes: $\Fwd{\dilateEndpoints}(\shape)$ (the result of the half-dilation of each endpoint) and $\Fwd{\dilateSegments}(\shape)$, where each segment $\seed_i$ is extruded along the dilation axis by its radius $\dilationRadius_i$.
This decomposition is illustrated in \Cref{fig:segments-separate}.

\begin{figure}[tbp!]
	\centering
	\includegraphics[width=0.75\linewidth]{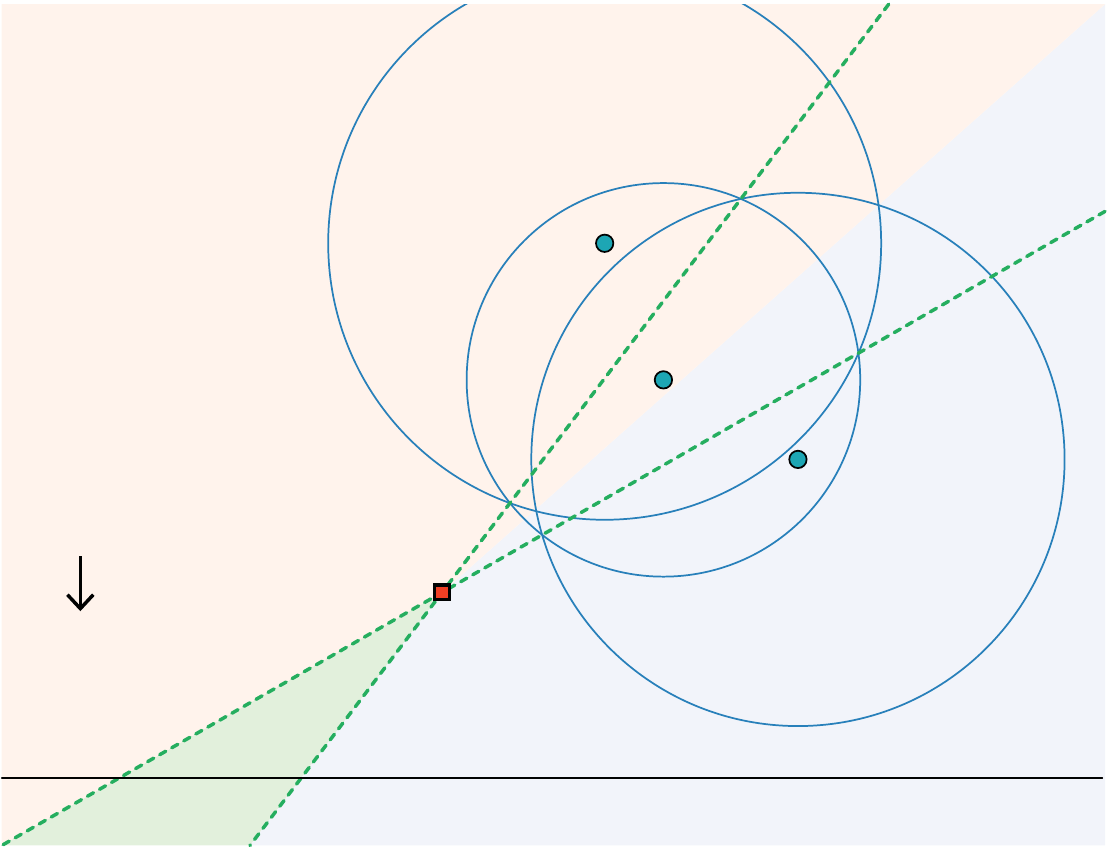}

	\caption{
		Sweepline algorithm computing $\protect\Fwd{\dilateEndpoints}(\shape)$.
		When inserting the rightmost point in the set of active seeds $\mL$, it may happen that the power vertex with its two neighbors be located further away along the sweep direction.
		In this case, the middle point should not be removed from $\mL$, as its power cell will continue to intersect the sweepline.
	}
	\label{fig:power-vertex}
\end{figure}

Now, the dilated shape $\Fwd{\dilateSegments}(\shape)$ is easy to compute with a forward sweep, since there are no Voronoi diagrams or power diagrams involved: we simply remove occluded subsegments, keeping the ones with the largest radius, and removing a seed $\seed_i$ once its distance to the sweepline is $> \dilationRadius_i$.
To compute $\Fwd{\dilateEndpoints}(\shape)$, we employ a sweep similar to the one described \Cref{subsec:method:voronoi-2d}, but now all the seeds are \emph{points}, which greatly simplifies the calculation of power vertices.
Indeed, the power bisectors are now lines, and the power cells are intersections of half-spaces (and thus convex).
The derivations for the coordinates of a power vertex is given in \Cref{app:power-vertex}.
The only special case to consider here is illustrated in \Cref{fig:power-vertex}: when inserting a new seed and updating $\mL$, the events corresponding to a power vertex do not always correspond to a removal. For example, in the situation illustrated in \Cref{fig:power-vertex}, the power cell of the middle point will continue to intersect the sweepline after it has passed the power vertex, so we should not remove the middle seed from $\mL$. Fortunately, this case is easy to detect, and we simply forgo the insertion of the event in $\mQ$.

\subsection{Complexity Analysis}
\label{subsec:complexity-analysis}

We analyze the runtime complexity of our algorithm, assuming that the dexel spacing is $1$, so that the dilation radius $\dilationRadius{}$ is given in the same unit as the dexel numbers.

For the first 2D dilation operation, the complexity of combining two dexel data structures is linear in the size of the input, since the segments are already sorted. The cost of the forward dilation operation $\Fwd{\dilate}_\dilationRadius(\shape)$ requires a more detailed analysis: Let $n$ be the number of input segments, and $m$ be the number of output segments in the dilated shape $\Fwd{\dilate}(\shape)$.
In the worst case, each input segment generates $\O(r)$ distinct output segments, $m = \O(nr)$, and each seed segment is split twice by every newly inserted segment. Since each seed can split at most one element of $\mL$ into two separate segments, we have that, at any time, $\size{\mL} = \O(n)$. Moreover, each seed produces at most three events in $\mQ$ (a Voronoi vertex with its left/right neighbors, and the moment it becomes inactive because of its distance to the sweepline). It follows that $\size{\mQ} = \O(n)$ as well.
Segments in $\mL$ are stored in a \inlinecode{std::set<>}, thus insertion and removal (\lrefs{alg:voronoi:sweep-line:remove-seed}{alg:voronoi:sweep-line:insert-seed} in \Cref{alg:voronoi:sweep-line}) can be performed in $\O(\log n)$ time.
While the line dilation (\lref{alg:voronoi:sweep-line:dilate-line} in \Cref{alg:voronoi:sweep-line}) is linear in the size of $\mL$, the total number of segments produced by this line cannot exceed $m$, so the amortized time complexity over the whole sweep is $\O(m)$.
This brings the final cost of the whole dilation algorithm to a time complexity that is $\O(n \log(n) + m)$, and it does not depend on the dilation radius $\dilationRadius{}$ (apart from the output size $m$). In contrast, the offsetting algorithm presented in~\cite{Wang:2013:GBO} has a total complexity that grows proportionally to $\dilationRadius{}^2$.

For the second dilation operation, where each input segment is associated a specific dilation radius, the result is similar. Indeed, $\Fwd{\dilateEndpoints}(\shape)$ is computed using the same algorithm as before, so the analysis still holds.
Combining the dexels in $\Fwd{\dilateEndpoints}(\shape)$ with the results from $\Fwd{\dilateSegments}(\shape)$ can be done linearly in the size of the output as we advance the sweepline, so the total complexity of computing $\Fwd{\dilatePower}(\shape)$ is still $\O(n \log(n) + m)$.

For the 3D case, the result of a first extrusion is used as input for the second stage dilation, the total complexity is more difficult to analyze, as it also depends on the structure of the intermediate result.
In a conservative estimate, bounding the number of intermediate segments by $\O(n\dilationRadius)$, this bring the final complexity of the 3D dilation to $\O(n\dilationRadius\log(n \dilationRadius{}) + m)$, where $m = \O(n\dilationRadius^2)$ is the size of the output model.
In practice, many segments can be merged in the final output, especially when the dilation radius is large, and $m$ may even be smaller than $n$ (e.g., when details are erased from the surface).

Finally, we note that in each stage of the 3D pipeline, the 2D dilations can be performed completely independently in every slice of the dexel structure, making the process trivial to parallelize.
We discuss the experimental performance of a multi-threaded implementation of our method in \Cref{sec:results}.

\section{Results}
\label{sec:results}
We implemented our algorithm in C++ using \rev{Eigen~\cite{Guennebaud:2010:EIG}} for linear algebra routines, and \rev{Intel Threading Building Blocks~\cite{Reinders:2010:ITB}} for parallelization. We ran our experiments on a desktop with a 6-core Intel\textsuperscript{\textregistered} Core\texttrademark{} i7{-}5930{K} CPU clocked at 3.5~GHz and 64 GB of memory. \rev{Our reference implementation is available on \fnurl{https://github.com/geometryprocessing/voroffset}{github} to simplify the adoption of our technique}.
\rev{Note that our results are sensitive for the choice of the dexel direction, since it will lead to different discretizations. In our experiments, we manually select this direction. For 3D printing applications, \textcite{Livesu:2017:FDM} [Section~3.2] give an overview of algorithms computing an optimized direction to increase the fidelity of the printing process.}

\begin{figure}[htbp!]
	\centering

	\makebox[\linewidth][c]{
	\includegraphics[scale=0.57]{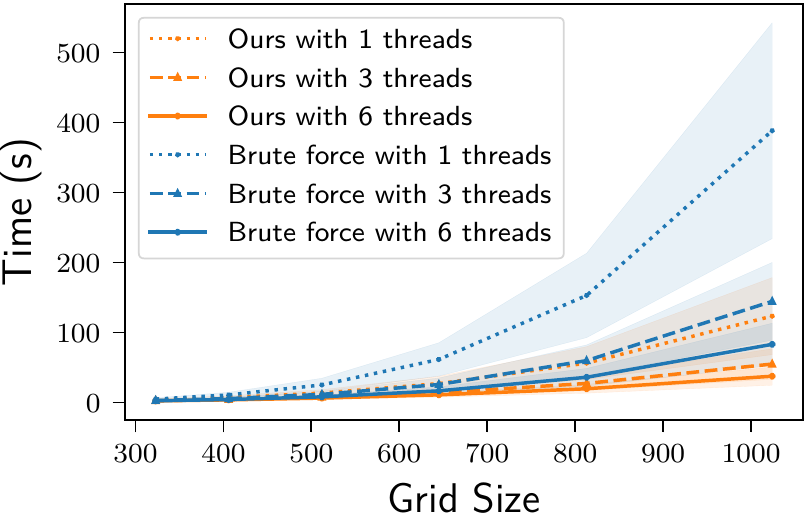}
	\includegraphics[scale=0.57]{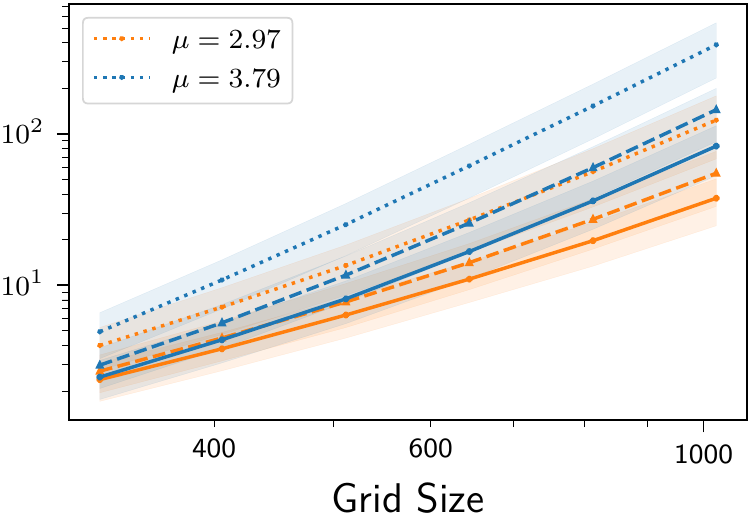}
	}

	\vspace{1ex}

	\makebox[\linewidth][c]{
	\includegraphics[scale=0.57]{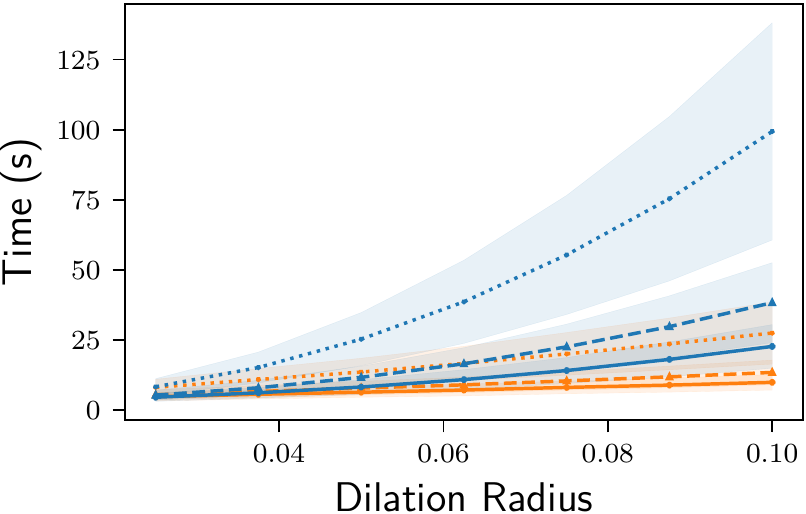}
	\includegraphics[scale=0.57]{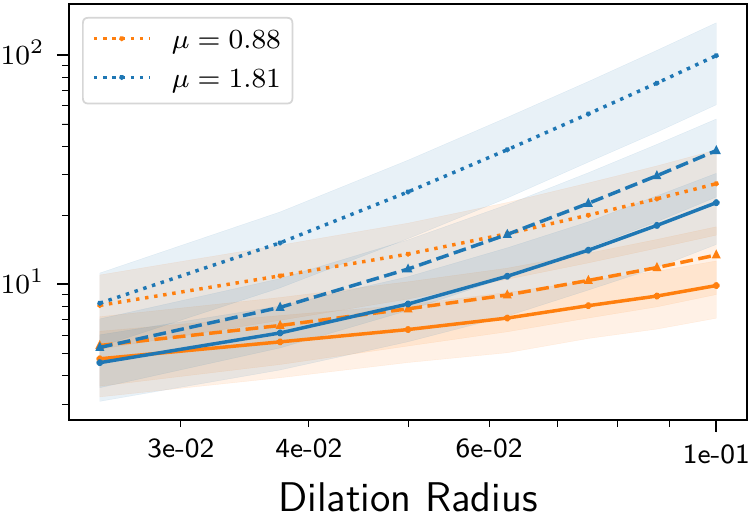}
	}

	\caption{
		Total running time averaged on 11 testing models, compared with a direct brute-force implementation.
		Standard deviation is shown in overlay, and convergence rate $\mu$ are reported in the log-log plots.
		Radii are relative to the grid size. The top row uses a relative radius of $0.05$, and the bottom row uses a grid size of $512^2$.
	}
	\label{fig:plot:timings}
\end{figure}

\paragraph{Baseline Comparison}
We implemented a simple brute-force dilation algorithm (on the dexel grid) to verify the correctness of our implementation, and to demonstrate the benefits of our technique.
In the brute-force algorithm, each segment in the input dexel structure generates an explicit list of dilated segments in a disk of radius $\dilationRadius{}$ around it, and all overlapping segments are merged in the output data structure.
\Cref{fig:plot:timings} compares the two methods using a different number of threads, and with respect to both grid size and dilation radius. In all cases, our algorithm is, as expected, superior not only asymptotically but also for a fixed grid size or dilation radius. Since each slice can be treated independently in our two-stage dilation process, our algorithm is embarrassingly parallel, and scales almost linearly with the number of threads used.

We note that the asymptotic time complexity observed in \Cref{fig:plot:timings} agrees with our analysis in \Cref{subsec:complexity-analysis}. Indeed, the dexel grid has a number of dexel $n \propto s^2$ is proportional to the squared grid size $s^2$, while the (absolute) dilation radius $\dilationRadius{}_{\mathrm{abs}} \propto s \dilationRadius{}_{\mathrm{rel}}$ grows linearly with the grid size. Since the complexity analysis in \Cref{subsec:complexity-analysis} uses a dilation radius $\dilationRadius{}$ expressed in dexel units, the observed asymptotic rate of $\approx 3$ indicates that our method is indeed $\O(s^3)$.

\begin{figure}[htbp!]
	\centering

	\makebox[\linewidth][b]{
	\includegraphics[scale=0.56]{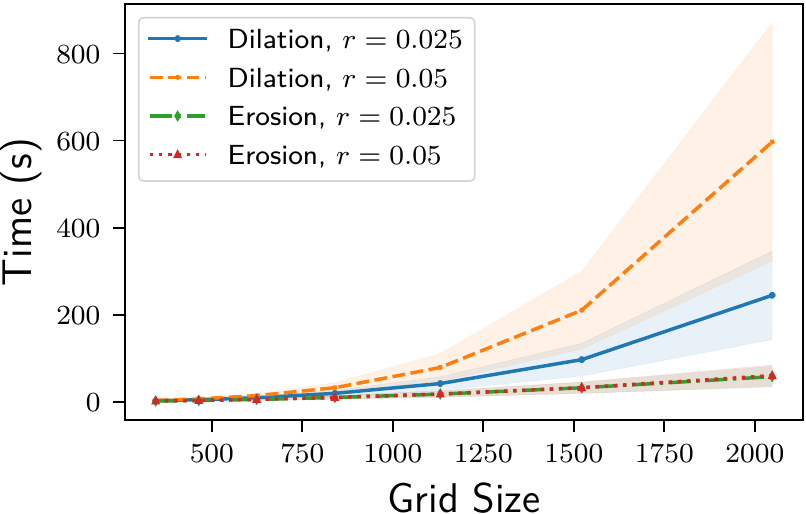}
	\includegraphics[scale=0.56]{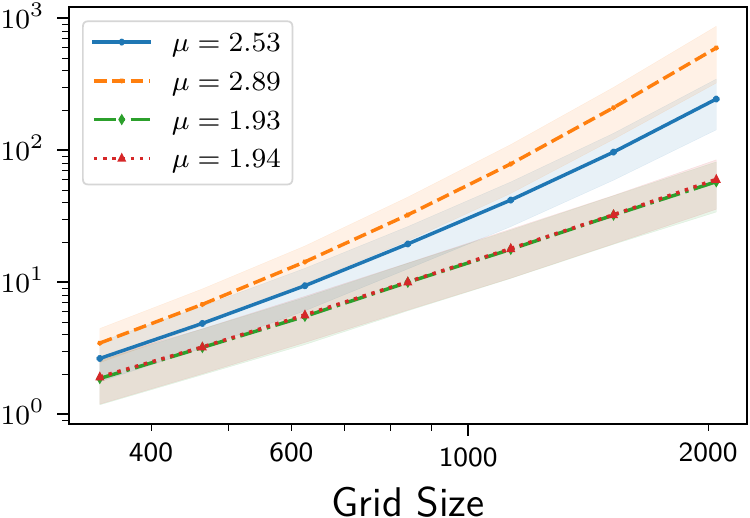}
	}

	\caption{
		Total running time for the dilation and erosion operators, for two different radii.
		Standard deviation is shown in overlay and the convergence rate $\mu$ is reported in a log-log plot.
	}
	\label{fig:plot:dilation-erosion}
\end{figure}

\paragraph{Dilation and Erosion}
In \Cref{fig:plot:dilation-erosion}, we compare the performance of the dilation and erosion operator on a small data set of 11 models, provided in the supplemental material. The dilation operator has a higher cost than the erosion, both asymptotically and in absolute running time. This is because the erosion operator reduces the number of dexels, leading to a considerable speedup.

\begin{figure*}[tbp!]
	\centering

	\sisetup{
		round-mode=figures,
		round-precision=3,
		scientific-notation=true,
		output-exponent-marker=\text{e}
	}

	\makebox[\linewidth][c]{
	\hspace*{-2em}
	\begin{tabular}{@{}c@{}c@{~}c@{~}c@{~}c@{~}c@{~}c@{~}c@{}}
		Model & &
		\multicolumn{2}{c}{\textcite{Campen:2010:PBE}} &
		\multicolumn{2}{c}{\textcite{Wang:2013:GBO}} &
		\multicolumn{2}{c}{Ours} \\
		\toprule

		\multirow{2}{*}{\mcenter{\includegraphics[height=3.5cm]{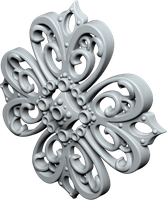}}} &
		$r_1$ &

		\mcenter{\includegraphics[height=3.5cm]{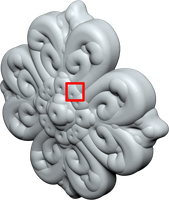}} &
		\raisebox{-1ex}{\specialcell{\includegraphics[height=2cm]{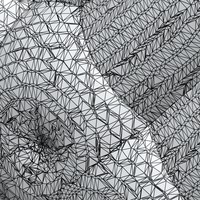} \\ $d_H$ = \num{0.184975}}} &
		\mcenter{\includegraphics[height=3.5cm]{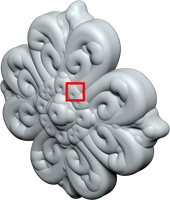}} &
		\raisebox{-1ex}{\specialcell{\includegraphics[height=2cm]{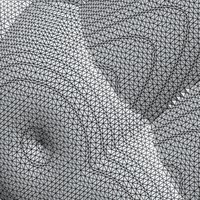} \\ $d_H$ = \num{0.16009}}} &
		\mcenter{\includegraphics[height=3.5cm]{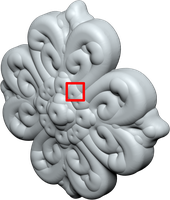}} &
		\raisebox{-1ex}{\specialcell{\includegraphics[height=2cm]{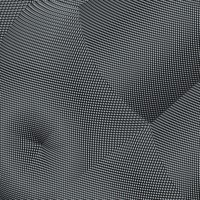} \\ ~}} \\
		\addlinespace[0.5ex]

		& $r_2$ &

		\mcenter{\includegraphics[height=3.5cm]{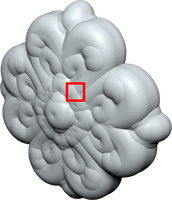}} &
		\raisebox{-1ex}{\specialcell{\includegraphics[height=2cm]{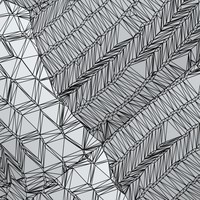} \\ $d_H$ = \num{0.160938}}} &
		\mcenter{\includegraphics[height=3.5cm]{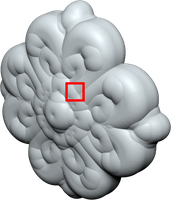}} &
		\raisebox{-1ex}{\specialcell{\includegraphics[height=2cm]{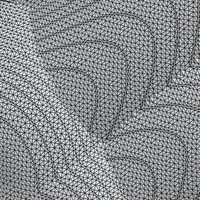} \\ $d_H$ = \num{0.146115}}} &
		\mcenter{\includegraphics[height=3.5cm]{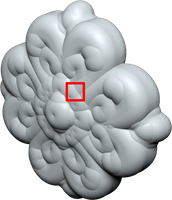}} &
		\raisebox{-1ex}{\specialcell{\includegraphics[height=2cm]{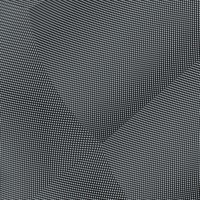} \\ ~}} \\

		\addlinespace
		\multirow{2}{*}{\mcenter{\includegraphics[height=3.5cm]{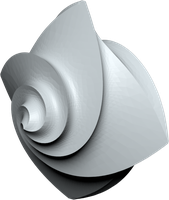}}} &
		$r_1$ &

		\mcenter{\includegraphics[height=3.5cm]{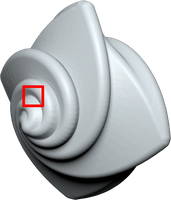}} &
		\raisebox{-1ex}{\specialcell{\includegraphics[height=2cm]{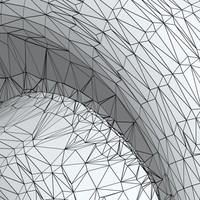} \\ $d_H$ = \num{0.0747678}}} &
		\mcenter{\includegraphics[height=3.5cm]{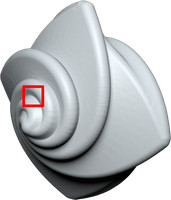}} &
		\raisebox{-1ex}{\specialcell{\includegraphics[height=2cm]{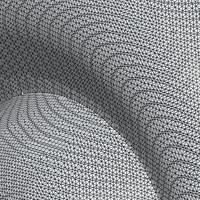} \\ $d_H$ = \num{0.0782019}}} &
		\mcenter{\includegraphics[height=3.5cm]{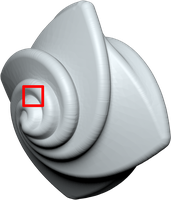}} &
		\raisebox{-1ex}{\specialcell{\includegraphics[height=2cm]{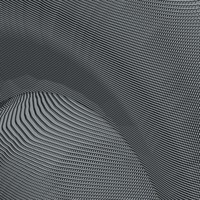} \\ ~}} \\
		\addlinespace[0.5ex]

		& $r_2$ &

		\mcenter{\includegraphics[height=3.5cm]{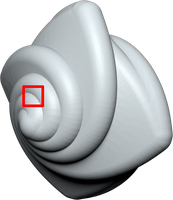}} &
		\raisebox{-1ex}{\specialcell{\includegraphics[height=2cm]{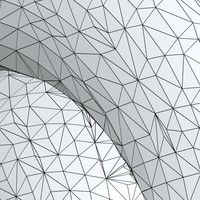} \\ $d_H$ = \num{0.0882963}}} &
		\mcenter{\includegraphics[height=3.5cm]{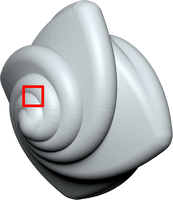}} &
		\raisebox{-1ex}{\specialcell{\includegraphics[height=2cm]{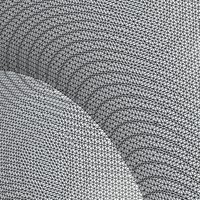} \\ $d_H$ = \num{0.0950772}}} &
		\mcenter{\includegraphics[height=3.5cm]{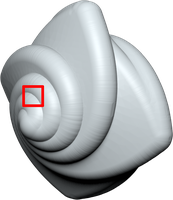}} &
		\raisebox{-1ex}{\specialcell{\includegraphics[height=2cm]{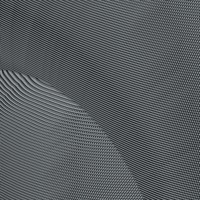} \\ ~}} \\

		\addlinespace
		\multirow{2}{*}{\mcenter{\includegraphics[height=3.5cm]{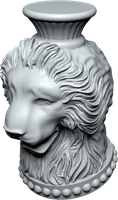}}} &
		$r_1$ &

		\mcenter{\includegraphics[height=3.5cm]{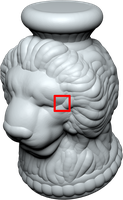}} &
		\raisebox{-1ex}{\specialcell{\includegraphics[height=2cm]{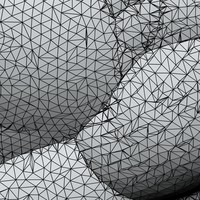} \\ $d_H$ = \num{0.155649}}} &
		\multicolumn{2}{c}{N/A} &
		\mcenter{\includegraphics[height=3.5cm]{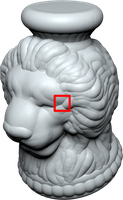}} &
		\raisebox{-1ex}{\specialcell{\includegraphics[height=2cm]{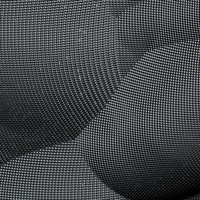} \\ ~}} \\
		\addlinespace[0.5ex]

		& $r_2$ &

		\mcenter{\includegraphics[height=3.5cm]{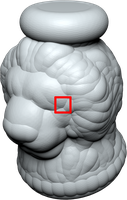}} &
		\raisebox{-1ex}{\specialcell{\includegraphics[height=2cm]{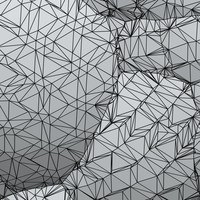} \\ $d_H$ = \num{0.11503}}} &
		\multicolumn{2}{c}{N/A} &
		\mcenter{\includegraphics[height=3.5cm]{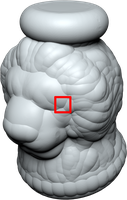}} &
		\raisebox{-1ex}{\specialcell{\includegraphics[height=2cm]{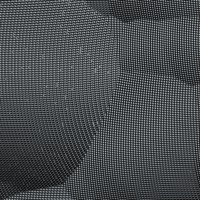} \\ ~}}
	\end{tabular}%
	}

	\caption{
		Quality comparison between~\cite{Campen:2010:PBE,Wang:2013:GBO}, and our method.
		The dilation radii $r_1$ and $r_2$ are set to $0.025 d$ and $0.05 d$ respectively, where $d$ is the bounding box diagonal of each model.
		The zoom insert corresponds to the red frame \rev{on each picture}. Both~\cite{Wang:2013:GBO} and our method use a discretization of $1024$ dexels.
		\rev{The Hausdorff distance $d_H$ between each result and ours is shown under the zoomed pictures, and is expressed as a \emph{percentage of the bounding box diagonal}.
		Note that the error corresponds roughly to the dexel size used for the discretization.}
		The surface from~\cite{Campen:2010:PBE} is reconstructed by marching on cells of an octree, and~\cite{Wang:2013:GBO} uses dual-contouring with normal information.
		For shading purposes, we show in our full-view a surface reconstructed from the dexel samples using~\cite{Boltcheva:2017:SRB}, while our zoomed view shows the raw dexels.
		Results for the vase using~\cite{Wang:2013:GBO} is unavailable due to a crash in the reference implementation provided by the authors.
	}
	\label{fig:compare-quality}
\end{figure*}

\paragraph{Comparison with \textcite{Wang:2013:GBO}}
The most closely related work on offset from ray-reps representations is \textcite{Wang:2013:GBO}. It proposes to perform an offset from a LDI sampled from three orthogonal directions. The offset is computed as the union of spheres sampled at the endpoints of each segment from all three directions at once. In contrast, our method relies on a single dexel structure, which has both advantages and drawbacks. It is applicable in a situation where only one view is available, or where one view is enough to describe the model (e.g., modeling for additive manufacturing~\cite{Lefebvre:2013:IAG}). The drawback is that it is less precise on the orthogonal directions (where the 3-views LDI will have more precise samples). However, the difference are minor, as shown in the comparison in \Cref{fig:compare-quality}.

We report a performance comparison in \Cref{fig:timings}, where we compared our 6-core CPU version (running on a Intel\textsuperscript{\textregistered} Core\texttrademark{} i7{-}5930{K}) with their GPU implementation~\cite{Wang:2013:GBO} (running on a GTX 1060). The timings of the two implementations are comparable, suggesting that our CPU implementation is competitive even with their GPU implementation. Their implementation runs out of memory for the larger resolution (\rev{$2048^3$}), while our implementation successfully computes the dilation.

Extending our method to the GPU is a challenging and notable venue for future work, that could enable real-time offsetting on large and complex dexel structures.

\begin{figure}[htbp!]
	\centering

	\sisetup{
		round-mode=places,
		round-precision=2
	}

	\makebox[\linewidth][c]{
	\begin{tabular}{c@{}cS[table-format=3.0]SSS}
	 & \multirow{2}{*}{Radius} & {\multirow{2}{*}{Resolution}} & \multicolumn{2}{c}{\textcite{Wang:2013:GBO}} & {\multirow{2}{*}{Ours}} \\
	 &        &            & {Normal} & {Successive} &  \\ \toprule
	\multirow{6}{*}{\rotatebox{90}{\centering Filigree}}
	 &       &  512 & 4.94265 & 2.24845 &   1.26 \\
	 & $r_1$ & 1024 & 73.8496 &  12.548 &   9.96 \\
	 &       & 2048 &     {-} &     {-} &  88.97 \\ \addlinespace
	 &       &  512 & 10.9648 & 3.64415 &   2.14 \\
	 & $r_2$ & 1024 & 176.652 & 22.9433 &  19.04 \\
	 &       & 2048 &     {-} &     {-} & 280.96 \\
	\midrule
	\multirow{6}{*}{\rotatebox{90}{\centering Octa-flower}}
	 &       &  512 & 5.93187 & 1.19794 &    2.39 \\
	 & $r_1$ & 1024 & 90.1667 & 8.90124 &   22.41 \\
	 &       & 2048 &     {-} &     {-} &  322.34 \\ \addlinespace
	 &       &  512 & 13.6807 & 2.53109 &    5.04 \\
	 & $r_2$ & 1024 & 224.403 & 19.5172 &   54.99 \\
	 &       & 2048 &     {-} &     {-} & 1649.56 \\
	\midrule
	\multirow{6}{*}{\rotatebox{90}{\centering Vase}}
	 &       &  512 &     {-} &      {-} &   1.38 \\
	 & $r_1$ & 1024 &     {-} &      {-} &  12.24 \\
	 &       & 2048 &     {-} &      {-} & 131.13 \\ \addlinespace
	 &       &  512 & 15.7145 &      {-} &   2.57 \\
	 & $r_2$ & 1024 &     {-} &      {-} &  25.61 \\
	 &       & 2048 &     {-} &      {-} &  524.9 \\
	\end{tabular}
	}
	\vspace{1ex}

 	\caption{%
 		Timing (\si{\second}) comparisons with~\cite{Wang:2013:GBO} across different dexel resolutions.
 		Dilation radii $r_1$ and $r_2$ are set to $0.025 d$ and $0.05 d$ respectively, where $d$ is the diagonal of bounding box of the model.
 		The two columns for~\cite{Wang:2013:GBO} corresponds to the \enquote{GPU Primary} and \enquote{GPU SH+P+Succ} in their Table~2 respectively.
 		A \enquote{-} indicates that the program terminated with an error (crash or went out of memory).
 		\cite{Wang:2013:GBO} ran on a GeForce GTX 1060, while our comparisons on a 6-core Intel\textsuperscript{\textregistered} Core\texttrademark{} i7{-}5930{K} CPU. Our multi-thread CPU implementation is competitive with a GPU implementation, while scaling to higher resolutions thanks to the memory efficient dexel data structure.%
 	}

	\label{fig:timings}
\end{figure}

\paragraph{Comparison with \textcite{Campen:2010:PBE}}
In \Cref{fig:compare-quality}, we compare our dilation algorithm (based on a dexel data structure) and~\cite{Campen:2010:PBE} (based on an octree), matching their parameters to get a similar final accuracy. Their running time is higher and depends on the input complexity (\SI{45}{\second} for Octa-flower, \SI{519}{\second} for Vase, and \SI{2294}{\second} for Filigree).
\rev{We computed the Hausdorff distance between our result and theirs (\Cref{fig:compare-quality}). In all cases the error remains in the order of the voxel/dexel size used for the discretization.
Our results are visually indistinguishable from theirs, but are computed at a small fraction of the cost.}

\begin{figure*}[htbp!]
	\centering
	\captionsetup[subfigure]{labelformat=empty}

	\makebox[\linewidth][c]{
	\subfloat[Input]{\includegraphics[width=0.2\linewidth]{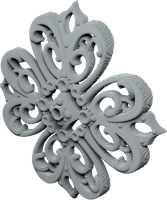}}
	\makebox[0.22\linewidth][c]{\subfloat[Dilation]{%
		\begin{overpic}[width=0.2\linewidth]{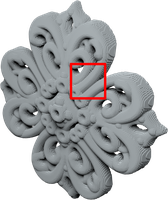}
			\put(60,65) {\frame{\includegraphics[width=0.08\linewidth]{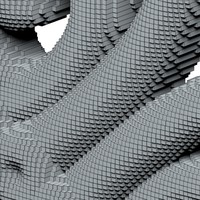}}}
		\end{overpic}%
	}}
	\makebox[0.22\linewidth][c]{\subfloat[Erosion]{%
		\begin{overpic}[width=0.2\linewidth]{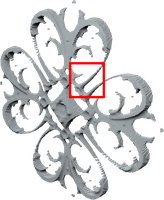}
			\put(59,65) {\frame{\includegraphics[width=0.08\linewidth]{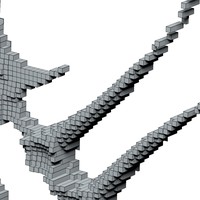}}}
		\end{overpic}%
	}}
	\makebox[0.22\linewidth][c]{\subfloat[Opening]{%
		\begin{overpic}[width=0.2\linewidth]{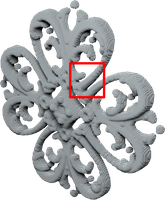}
			\put(59,65) {\frame{\includegraphics[width=0.08\linewidth]{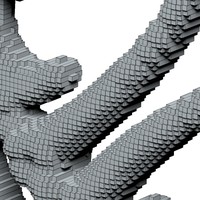}}}
		\end{overpic}%
	}}
	\makebox[0.22\linewidth][c]{\subfloat[Closing]{%
		\begin{overpic}[width=0.2\linewidth]{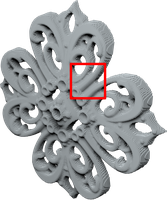}
			\put(59,65) {\frame{\includegraphics[width=0.08\linewidth]{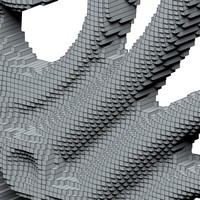}}}
		\end{overpic}%
	}}
	}
	\caption{
		Topological cleaning of an input shape via morphological operations, using a grid of $256 \times 256$ dexels.
		A hires zoomed-in view of the framed area is shown atop each result.
	}
    \label{fig:cleaning}
\end{figure*}

\paragraph{Topological Cleaning}
Our efficient dilation and erosion operators can be combined to obtain efficient opening and closing operators (\Cref{fig:cleaning}). For example, the closing operation can be used to remove topological noise, i.e., small handles, by first dilating the shape by a fixed offset, and then partially undoing it using erosion. While most regions of the object will recover their original shape, small holes and sharp features will not, providing an effective way to simplify the  topology.

\begin{figure*}[htbp!]
	\centering
	\captionsetup[subfigure]{labelformat=empty,justification=centering}

	\makebox[\linewidth][c]{
	\subfloat[Input Surface]{\includegraphics[height=5cm]{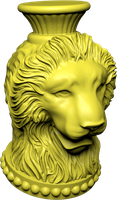}}
	\hfill{}
    \subfloat[][Grid Size $\mathsf{128^2}$ \\ \SI{0.06}{\second}]{\includegraphics[height=5cm]{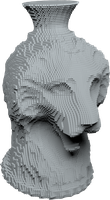}}
	\hfill{}
	\subfloat[][Grid Size $\mathsf{256^2}$ \\ \SI{0.23}{\second}]{\includegraphics[height=5cm]{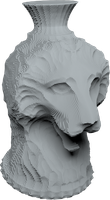}}
	\hfill{}
	\subfloat[][Grid Size $\mathsf{512^2}$ \\ \SI{1.7}{\second}]{\includegraphics[height=5cm]{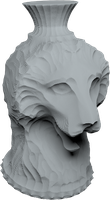}}
	\hfill{}
	\subfloat[][Grid Size $\mathsf{1024^2}$ \\ \SI{12.03}{\second}]{\includegraphics[height=5cm]{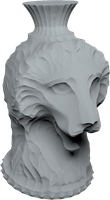}}
	\hfill{}
	\subfloat[][Grid Size $\mathsf{2048^2}$ \\ \SI{110.98}{\second}]{\includegraphics[height=5cm]{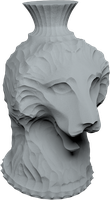}}
	}
	\caption{
		Result of an erosion operation using different grid resolutions.
	}
    \label{fig:resolution}
\end{figure*}

\paragraph{Scalability}
The compactness of the dexel representation enables us to represent and process immense volumes on normal desktop computers. An example is shown in \Cref{fig:resolution} for the erosion operation. Note that the results on the right have a resolution sufficiently high to hide the dexels: this resolution would be prohibitive with a traditional boundary or voxel representation (\Cref{fig:timings}).

\paragraph{3D Printing}
The Boolean difference between a dilation and erosion of a shape produces a shell of controllable thickness. This operation is useful in 3D printing applications, since the interior of an object is usually left void (or filled with support structures) to save material.
Another typical use case for creating thick shells out of a surface mesh is the creation of molds~\cite{Malomo:2016:FAD}. We show a high resolution example of this procedure in \Cref{fig:teaser}, and we fabricated the computed shell using PLA plastic on an Ultimaker 3 printer.

\paragraph{Limitations}
The main limitation of our method comes from the uneven sampling of the dexel data structure in a single direction (e.g., the flat areas on the sides of the reconstructed models in \Cref{fig:compare-quality}). While this is not a problem if the application only needs a certain resolution to begin with (e.g., 3D printing or CT scans), it is not optimal for the purpose of reconstructing an output mesh, where other approaches such as~\cite{Campen:2010:PBE,Wang:2013:GBO} will lead to a higher fidelity. We could use our method with a dexelized structure in 3 orthogonal directions (similar to LDI~\cite{Huang:2014:AFL}), and reconstruct the output surface using~\cite{Boltcheva:2017:SRB}, but this will lead to a threefold increase of the running times.

\section{Future Work and Concluding Remarks}
\label{sec:conclusion}

Our current algorithm is restricted to uniform morphological operations. It would be worthwhile to extend it to single direction thickening (for example in the normal direction only) or to directly work on a LDI offset, i.e., representing the shape with 3 dexel representations, one for each axis. A GPU implementation of our technique would likely provide a sufficient speedup to enable real-time processing of ray-reps representation at the resolution typically used by modern 3D printers.

We proposed an algorithm to efficiently compute morphological operations on ray-rep representations, targeting in particular the generation of surface offsets.
Beside offering theoretical insights on power diagrams and their application to surface offsets, our algorithm is simple, robust, and efficient: it is an ideal tool in 3D printing applications, since it can directly process voxel or dexel representations to filter out topological noise or extract volumetric shells from boundary representations.

\section*{Acknowledgments}

This work was supported in part by the NSF CAREER award IIS-1652515, the NSF grant OAC-1835712, a gift from Adobe, and a gift from nTopology.

\printbibliography

\begin{IEEEbiography}[{\vspace*{-0.125in}\includegraphics[width=1in,height=1.25in,clip,keepaspectratio]{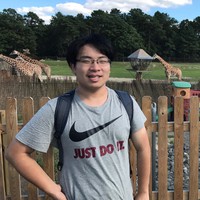}}]{Zhen Chen}
is a first year PhD student at the Department of Computer Science in the University of Texas at Austin. He earned his Bachelor degree in Computational Mathematics at the University of Science and Technology of China in 2018.
From June to August 2017, he has been a visiting student at the Courant Institute of Mathematical Sciences (New York University, USA)\@. His research interests are 3D printing, geometry processing and shell deformation.
\end{IEEEbiography}

\begin{IEEEbiography}[{\includegraphics[width=1in,height=1.25in,clip,keepaspectratio]{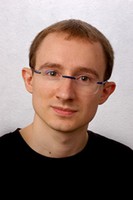}}]{Daniele Panozzo}
Daniele Panozzo is an assistant professor of computer science at the Courant Institute of Mathematical Sciences in New York University. Prior to joining NYU he was a postdoctoral researcher at ETH Zurich (2012--2015). He earned his PhD in Computer Science from the University of Genova (2012) and his doctoral thesis received the EUROGRAPHICS Award for Best PhD Thesis (2013). He received the EUROGRAPHICS Young Researcher Award in 2015 and the NSF CAREER Award in 2017. Daniele is leading the development of libigl (\url{https://github.com/libigl/libigl}), an award-winning (EUROGRAPHICS Symposium of Geometry Processing Software Award, 2015) open-source geometry processing library that supports academic and industrial research and practice. Daniele is chairing the Graphics Replicability Stamp (\url{http://www.replicabilitystamp.org}), which is an initiative to promote reproducibility of research results and to allow scientists and practitioners to immediately benefit from state-of-the-art research results. Daniele's research interests are in digital fabrication, geometry processing, architectural geometry, and discrete differential geometry.
\end{IEEEbiography}

\begin{IEEEbiography}[{\includegraphics[width=1in,height=1in,clip,keepaspectratio]{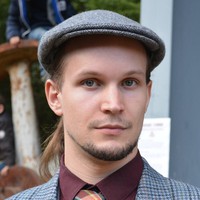}}]{J\'{e}r\'{e}mie Dumas}
J\'{e}r\'{e}mie Dumas is a research engineer at nTopology Inc.\ in New York. Prior to joining nTopology Inc.\, he was a postdoctoral fellow at the Courant Institute of Mathematical Sciences in New York University.
J\'{e}r\'{e}mie completed his PhD at INRIA Nancy Grand-Est in 2017, under the direction of Sylvain Lefebvre.
His doctoral thesis received the EUROGRAPHICS Award for Best PhD Thesis (2018). His work focuses on shape synthesis for digital fabrication, shape optimization, simulation, microstructures, and procedural synthesis.
\end{IEEEbiography}

\vfill

\newpage

\appendices
\crefalias{section}{appsec}

\section{Voronoi Vertex between Three Parallel Segments in 2D}
\label{app:voronoi-vertex}

The bisector of two parallel segment seeds is 2D is a piecewise-quadratic curve, as illustrated in \Cref{fig:voronoi-vertex}.
A Voronoi vertex is a point at the intersection of the bisector curves between three segment seeds.
Because the segment seeds are non-overlapping, the Voronoi vertex can be either between three points (\Cref{app:voronoi-vertex-three-points}) or between one segment and two points (\Cref{app:voronoi-vertex-segment-points}).

\subsection{Voronoi Vertex between Three Points}
\label{app:voronoi-vertex-three-points}

Let $\point_1, \point_2, \point_3$ be three points, with coordinates $\point_i = (y_i, z_i)$.
A point $\point$ lying on the bisector of $(\point_1, \point_2)$ satisfies
\begin{equation}
\begin{aligned}
	\norm{\point - \point_1}^2 &= \norm{\point - \point_2}^2 \\
	\iff (y-y_1)^2+(z-z_1)^2 &= (y-y_2)^2+(z-z_2)^2
\end{aligned}
\end{equation}

After simplification, we get
\begin{equation}
	2(y_2-y_1)y+2(z_2-z_1)z=(y_2^2+z_2^2)-(y_1^2+z_1^2)
	\label{eq:vor-bisector-left-2pts}
\end{equation}

Similarly, for a point lying on the bisector of $(\point_2, \point_3)$,
\begin{equation}
	2(y_3-y_2)y+2(z_3-z_2)z=(y_3^2+z_3^2)-(y_2^2+z_2^2)
	\label{eq:vor-bisector-right-2pts}
\end{equation}

We can get the coordinate of Voronoi vertex between three points by solving the system formed by \Cref{eq:vor-bisector-left-2pts,eq:vor-bisector-right-2pts}:
\begin{equation}
	2 \Pmqty{
		y_2-y_1 & z_2-z_1 \\
		y_3-y_2 & z_3-z_2
	}
	\Pmqty{y \\ z} =
	\Pmqty{
		(y_2^2+z_2^2)-(y_1^2+z_1^2) \\
		(y_3^2+z_3^2)-(y_2^2+z_2^2)
	}
\end{equation}

\subsection{Voronoi Vertex between a Segment and Two Points}
\label{app:voronoi-vertex-segment-points}

Let $\point_1(y_1, z_1),\seed(y_s, z\sIn_s, z\sOut_s),\point_2(y_2, z_2)$ be three seeds.
We only need to consider the case where $y_s < y_1 < y_2$, other cases are similar.
For the Voronoi vertex $\point = (y,z)$ to intersect the bisectors where it is closest to the interior of $\seed$, and not one its endpoints, we need to have $z\sIn_s < z < z\sOut_s$.
It follows  the distance from $\point$ to the segment $\seed$ is:
\begin{align}
	\begin{aligned}
		\dist(\point, \seed) & = \inf\limits_{(a, b)\in \seed} (y - a)^2+(z - b)^2\\
		                 & = (y - y_s)^2 \quad \text{when~} (a, b) = (y_s, z)
	\end{aligned}
\end{align}

By definition of the Voronoi vertex,
\begin{align}
	\dist(\point, \seed) = \dist(\point, \point_1)
	\label{eq:vor-bisector-pt-seg}
	\\
	\dist(\point, \point_1) = \dist(\point, \point_2)
	\label{eq:vor-bisector-2pts}
\end{align}

Developing \Cref{eq:vor-bisector-pt-seg}, we get
\begin{equation}
\begin{aligned}
	(y-y_s)^2 &= (y-y_1)^2 + (z-z_1)^2 \\
	\iff 2(y_1-y_s)y-(z-z_1)^2 &= z_1^2-y_s^2
\end{aligned}
\end{equation}

Similarly, developing \Cref{eq:vor-bisector-2pts} leads to
\begin{align}
	2(y_1-y_2)y+2(z_1-z_2)z=(y_1^2+z_1^2)-(y_2^2+z_2^2)
\end{align}

Now, let
\begin{equation*}
	\left\{
	\begin{aligned}
		u & = 2(y_1-y_s) \\
		w & = -y_1^2+y_s^2 \\
		a & = 2(y_1-y_2) \\
		b & = 2(z_1-z_2) \\
		c & = (y_1^2+z_1^2)-(y_2^2+z_2^2) \\
	\end{aligned}
	\right.
\end{equation*}

We can rewrite the system of equations as
\begin{align}
	& \left\{
	\begin{aligned}
		uy-(z-z_1)^2+w &= 0 \\
		ay+bz          &= c \label{eq:vor-bisector-system-2}
	\end{aligned}
	\right.
	\\ \implies
	& az^2+(bu-2az_1)y+az_1-cu-aw=0
	\label{eq:vor-bisector-system}
\end{align}

Solving \Cref{eq:vor-bisector-system}, if the roots exist, we will have
\begin{align*}
	\Delta &= (bu-2a_1z)^2-4a(az_1^2-cu-aw) \\
	z_{1,2}^{*} &= \frac{2az_1-bu \pm \sqrt{\Delta}}{2a}
\end{align*}

Choosing the solution that belongs to $[z\sIn_s,z\sOut_s]$, and substituting into $ay+bz=c$, we will get the $y$-coordinate of our Voronoi vertex.

\section{Power Vertex between Three Points in 2D}
\label{app:power-vertex}

Let $\point_1(y_1,z_1;r_1),\point_2(y_2,z_2;r_2),\point_3(y_3,z_3;r_3)$ be three seeds. A power vertex can be computed from the intersection of two bisector lines.

A point $\point(y,z)$ lying on the bisector of $(\point_1,\point_2)$ satisfies:
\begin{equation}
\begin{alignedat}{4}
	 && \norm{\point - \point_1}^2 - r_1^2 &= \norm{\point - \point_2}^2 - r_2^2 \\
	\iff &&
		(y_1-y)^2+(z-z_1)^2-r_1^2 &= (y_2-y)^2+(z-z_2)^2-r_2^2 \\
	\iff &&
		2(y_2-y_1)y+2(z_2-z_1)z&=(y_2^2+z_2^2-r_2^2)-(y_1^2+z_1^2-r_1^2)
\end{alignedat}
\end{equation}

A similar equation holds for the bisector of $(\point_2, \point_3)$. This translates into the following system of equations:
\begin{align}
	2 \Pmqty{
		y_2-y_1 & z_2-z_1 \\
		y_3-y_2 & z_3-z_2
	}
	\Pmqty{y \\ z} =
	\Pmqty{
		(y_2^2+z_2^2-r_2^2)-(y_1^2+z_1^2-r_1^2) \\
		(y_3^2+z_3^2-r_3^2)-(y_2^2+z_2^2-r_2^2)
	}
\end{align}

\end{document}